\documentclass[natbib]{svjour3}                     % onecolumn (standard format)
\smartqed  % flush right qed marks, e.g. at end of proof
\usepackage{graphicx}
\newcommand{\la}{\,\rlap{\raise 0.5ex\hbox{$<$}}{\lower 1.0ex\hbox{$\sim$}}\,}
\newcommand{\ga}{\,\rlap{\raise 0.5ex\hbox{$>$}}{\lower 1.0ex\hbox{$\sim$}}\,}

\newcommand{{\z}}{{\.z}}
\newcommand{{\Z}}{{\.Z}}
\newcommand{{\rg}}{{r$_{\mathrm{g}}$}}

\def\suzaku{{\em Suzaku}\ }

\begin{document}

\title{X-ray Absorption and Reflection in Active Galactic Nuclei%\thanks{Grants or other notes
}


\author{T.J.Turner         \and
        L.Miller %etc.
}


\institute{T.J.Turner \at
              Dept. of Physics, University of Maryland Baltimore County, 1000 Hilltop Circle, Baltimore, MD 21250, USA\\
              Tel: +1-410-455-1978\\
              Fax: +1-410-455-1072\\
              \email{tjturner@umbc.edu}           %  \\
           \and
           L.Miller \at
              Dept. of Physics, University of Oxford, Denys Wilkinson Building, Keble Road, Oxford OX1 3RH, UK 
}

\date{Received: date / Accepted: date}

\maketitle

\begin{abstract}

X-ray spectroscopy offers an opportunity to study the complex mixture of emitting and
absorbing components in the circumnuclear regions of active galactic nuclei (AGN), and
to learn about the accretion process that fuels AGN 
and the feedback of material to 
their host galaxies.
We describe the spectral signatures that may be studied and 
review the X-ray spectra and spectral variability of active
galaxies, concentrating on progress from recent {\it Chandra}, {\it XMM-Newton} 
and {\it Suzaku} data for local type\,1 AGN. 
We describe the evidence for absorption covering a wide range of column densities, 
ionization and dynamics, and discuss the 
growing evidence for partial-covering absorption from data at energies $\ga 10$\,keV.
Such absorption
can also explain the observed X-ray spectral curvature and variability in AGN at lower energies
and is likely an 
important  factor in shaping the observed properties of this class of source. 
Consideration of self-consistent models for 
local AGN indicates that X-ray spectra likely comprise a combination of absorption and reflection effects 
from material originating within a  few light days  of the black hole as well as  on
larger scales.
It is likely that AGN X-ray spectra may be strongly affected by the presence of disk-wind outflows
that are expected in systems with high accretion rates, and we describe models that attempt to
predict the effects of radiative transfer through such winds, and discuss the prospects for
new data to test and address these ideas.

\keywords{ Galaxies: active \and X-rays: galaxies}  

\end{abstract}

\section{Introduction}
\label{sec:Intro}
Measurements of gas and stellar kinematics have shown that most galaxies 
harbor a black hole at their center 
\citep{kormendy95a,magorrian98a,ferrarese00a}.
The black holes in these galaxies are dubbed `supermassive',  
with masses in the range 
$10^6-10^9$\,M$_\odot$. The mass of the central black hole scales with galaxy bulge 
properties such as its velocity dispersion
\citep{gebhardt00a,tremaine02a}, its luminosity \citep{marconi03a} and its
stellar density profile \citep{graham01a},
and so the formation of the galaxy and the nuclear supermassive black hole must be linked. 
In active galactic nuclei (AGN), radiation  from 
the nuclear region is detected 
that is thought to be released from material accreting onto the black hole.
Studying AGN offers the opportunity to understand the coevolution of 
galaxies and black holes, as accretion must be linked to growth of the black hole
\citep{soltan82a}. The masses of black holes in active galaxies may also be estimated 
using  ``reverberation mapping'' 
\citep{peterson93a, kaspi00a, peterson00a, peterson04a}
giving results consistent with masses derived from stellar kinematics and luminosities
\citep{wandel02a, onken04a}  
and with the black hole masses required to explain the linewidths of the photoionized
broad-line regions in AGN \citep{wandel99a}.
The majority of nuclear black holes in the local Universe
are evident only through their gravitational influence and
either have no detectable signatures of accretion or are only weakly active
\citep[e.g.][]{heckman04a}.
The AGN phenomenon is observed over a wide range 
of intrinsic luminosity from dwarf Seyfert galaxies to 
quasars ($L_{\mathrm{bol}} \sim 10^{40}-10^{47} {\rm erg\ s^{-1}}$)
and shows strong cosmological evolution, with a greater density 
of the most luminous AGN at earlier cosmic epochs \citep[e.g.][]{boyle00a}.
However, the way in which black holes grow, and the conditions that cause
active nuclei to switch off, are still being debated. 

With the advent of X-ray observing satellites  it was  established \citep{elvis78a} that 
X-ray emission is a common property of active galaxies and that the  
X-ray flux comprises a significant fraction (about $5 - 40\%$: \citealt{ward87a}) 
of the bolometric emission from such objects. Observation of rapid (down to ks) variability  
in the X-ray flux of local Seyfert galaxies constrained the X-ray emission 
to be small and thus likely to arise 
very close to the active nucleus \citep[e.g.][]{lawrence85a,pounds86a} 
\footnote{
From the light travel-time across a source, in the absence of
relativistic effects, a characteristic timescale of 1\,ks corresponds to an  
upper limit to its size $D \la 200 M_6^{-1}$, expressed in units of the 
gravitational radius r$_{\mathrm g} = {\mathrm G}M/{\mathrm c}^2$,
where $M_6$ is the mass of a black hole in units of $10^6$\,M$_{\odot}$.}
The origin of X-rays from close to the central black hole means that 
X-ray data offer a chance to study the immediate environs of supermassive black holes and the 
poorly understood accretion process that fuels them. Although the X-ray emission 
region is too small to image with current instrumentation, timing analysis
 and spectroscopy offer ways to probe these regions indirectly. 
 In addition to tracing spectral signatures of 
the gas inflow and outflow at the heart of these accreting systems, X-ray data 
carry signatures of reprocessing in material within a few hundred gravitational radii and the 
potential for detecting measurable signatures from the accretion disk at even smaller radii.

\begin{figure}
\begin{center} 
\hbox{
\includegraphics[scale=0.35]{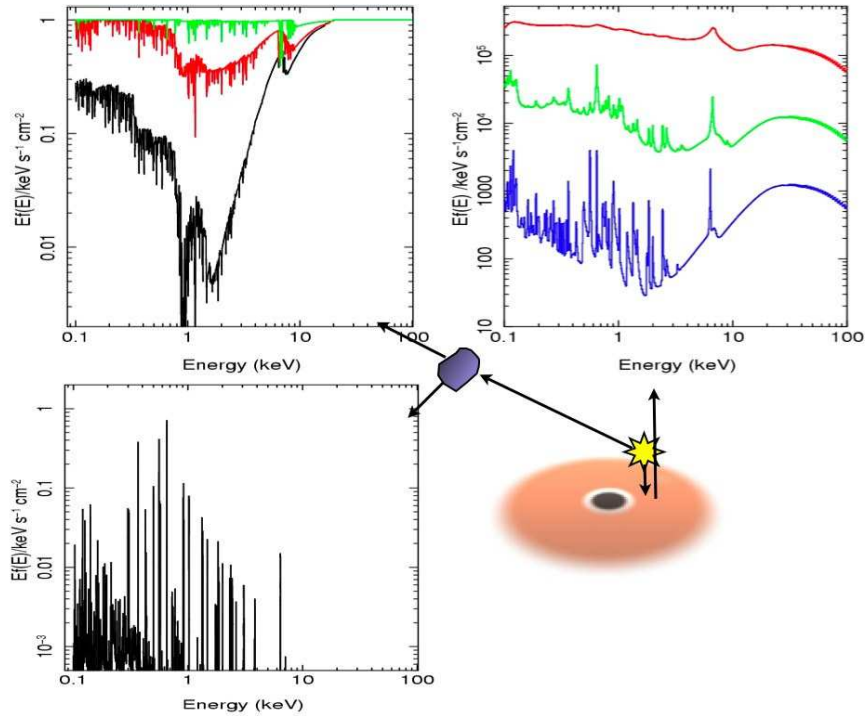}
}
\end{center}
\caption{\footnotesize
{A schematic of the accretion disk-black hole system. The region 
producing the primary X-ray continuum is denoted by a yellow star (this
may be a ``hotspot'' or series of hotspots or an extended coronal region).
The figure shows 
one sight-line where continuum reflection is seen and another 
where an absorption signature is imprinted onto the continuum. 
The top left panel shows the transmitted spectrum through gas with 
${\mathrm{N_H}} =5 \times 10^{23} {\rm cm^{-2}}$, log $\xi=2$(black), 3(red) and 4(green),
where $\xi$ is the ionization parameter in units of erg\,cm\,s$^{-1}$ 
(see section\,\ref{cthin}).
The absorption model was generated from {\sc xstar} 2.1ln8 with turbulent 
velocity $200$\,km\,s$^{-1}$ and density $n=10^{12} {\rm cm^{-3}}$. 
Line re-emission should also be observed from the absorbing clouds; 
the emission spectrum from the  log $\xi=2$ cloud is shown in the lower left panel.  
The top right panel shows reflection from the ionized surface of the accretion disk 
seen face-on. The model is from the {\sc reflionx} public model
table of \citet{ross05a} 
and shown for log $\xi=2$(blue), 3(green) and 4(red), shifted in normalization 
for clarity (and therefore in arbitrary units).  The primary continuum,
which in practice dilutes the reflection spectrum, is not shown.
Both models assumed solar abundances and a power-law 
continuum with photon index $\Gamma=2$.  
}} 
\label{fig:schematic}
\end{figure}

At optical wavelengths AGN are usually classified as either ``type\,1'', showing
broad optical/UV emission lines, or ``type\,2'', showing only narrow lines
\citep[see][]{osterbrock06a}.  The ``standard model'' supposes that in type\,2 AGN
the broad line region (BLR) exists but is highly extinguished, and it is usually supposed
that the extinguishing region forms a torus around the central region. 
\citep[e.g.][]{antonucci93a}.  
One expectation of the unified model, where the primary difference
between types is the orientation with respect to the observer, is that any phenomena
that are observed in type\,2 AGN should also be present in type\,1 AGN, perhaps modified
by a differing viewing angle.
This concept has been extended to the X-ray regime, where AGN whose X-ray 
emission is absorbed by intervening material are known as type\,2, and those without
significant absorption being evident are known as type\,1.  Although in general there is
often a good correspondence between optical and X-ray classifications
\citep[e.g.][]{bassani99a}, this agreement is by no means perfect \citep[e.g.][]{panessa02a}.

Within this broad classification there are intermediate types with detectable but weak
broad-line components on their optical/UV emission lines \citep{osterbrock06a},
designated by, e.g. ``type\,1.5''.
\citet{osterbrock85a} also identified a class of ``narrow-line Seyfert 1'' galaxies
that have many of the properties of type\,1 AGN but with rather narrow Balmer lines.
This class may comprise AGN accreting close to their Eddington limits
\citep{leighly99a, turner99a} and as such are particularly interesting if we wish to
study emission from accretion disks.  At the other end of the scale, 
a high fraction of nearby galaxies have ``Low Ionisation Nuclear Emission Regions''
(LINERs: \citealt{heckman80a}): 
the accretion in these objects may be in the form of a radiatively inefficient
flow \citep[e.g.][and see section~\ref{continuum}]{ho99a,ho08a}.

In this review we shall discuss physical processes at work in 
the X-ray regime and the progress that has been made over the last few decades. We will 
highlight the most recent observational results and critically address the degree to 
which we can constrain conditions in the environs of active nuclei using current 
X-ray instruments. As we are focusing on determining fundamentals of the accretion process 
we will limit this review to the properties of  bright nearby Seyfert type\,1 AGN for 
which the most detailed observational data are available, and whose timescales for variability 
are short enough for us to study the full range of their behavior. 
One of the original motivations for the unified scheme was to explain the radio properties
of radio-luminous AGN \citep{urry95a}.  It is known that radio-loud AGN are stronger
X-ray sources, with X-ray emission possibly being associated with a nuclear radio jet
\citep{zamorani81a} and in this review we shall discuss X-ray emission only from radio-quiet
AGN (although we should remember that all types of AGN have some level of nuclear radio
emission).

\section{X-ray continuum production and reprocessing}

\subsection{Continuum production}
\label{continuum}
Accretion onto the central black hole has become the accepted model for AGN,
as it is otherwise difficult to produce the observed radiation intensity over
such a wide range of the electromagnetic spectrum.
Accreting gas has angular momentum and we expect this material to form
an  accretion disk. When the gravitational potential 
energy lost by material moving inwards 
is released locally and efficiently through viscous dissipation, the 
gas becomes much cooler than the local  virial temperature and forms a 
disk whose vertical thickness is much smaller than the radius, the so-called 
 ``thin disk'' of \citet{shakura73a}. 
The disk emission is a composite of black-body radiation with a range of 
temperatures that depends on the range of emitting radii and the mass 
of the central black hole. 
For very low accretion rates an advection-dominated accretion flow (ADAF) 
may develop \citep[e.g.][]{narayan94a, abramowicz95a}, and as 
cooling by advection is inefficient the thermal energy in the disk is high  
and it becomes relatively thick in the vertical direction. 
These flows may be at work in low luminosity AGN such as LINERs. 

Constraints on accretion can be obtained from 
consideration of some basic observable quantities. The observed X-ray background radiation 
and the integrated QSO luminosity are consistent with the
expected integrated emission from the luminous accretion phase of AGN, given the estimated
local mass density of black holes, if 
around 10 percent of accreted rest energy has been radiated \citep{yu02a,marconi04a}.
This leads to an expectation
that most of the energy output from AGN occurs within a few gravitational radii of
the black hole \citep{fabian08a}, but we note here that this does not necessarily mean that 
the X-ray emission we {\em see} comes to us directly from close to the black hole. 
Whether or not we see any X-ray emission from the accretion disk at such small radii 
depends on the accretion disk extending close to the black hole and on there being
sufficiently concentrated illuminating radiation to generate significant reflection and
on there being no subsequent reprocessing of radiation by material further out. 
The estimate of this mean ``radiative efficiency'' from the X-ray background 
also depends critically on the range of redshifts 
over which the dominant growth occurs: this is constrained by the observed AGN 
luminosity function but we must be sure that the luminosity function in some waveband
adequately reflects the black hole growth at each epoch.

In the case of thin accretion disks around
supermassive black holes the intensity peak of the radiation is in the 
UV regime \citep{shakura73a}. 
To obtain X-ray emission, it is thought that UV photons from the 
innermost edge of the accretion disk serve as `seed photons' for  
multiple inverse Compton scatterings by a population of hot or relativistic 
electrons existing in a coronal region sandwiching the disk \citep{haardt91a,haardt93a,zdziarski94a}.
The origin of the Comptonizing electrons is at present uncertain but the coronal heating 
may be related to 
magnetic dissipation processes, or the corona may be populated with electron-positron 
pairs produced by 
photon-photon collisions (see, e.g., \citealt{fabian94a} and references therein). 

The Comptonizing electrons  
may have  a  thermal, non-thermal or mixed electron population. 
In this basic picture the observed X-ray 
continuum represents the  Comptonized spectrum whose shape depends 
upon the parameters of the electron population \citep[e.g.][]{haardt91a, titarchuk97a}. 
For example, the temperature of a thermal population can be parameterized as 
$\Theta_{\mathrm e}={\mathrm k}T_{\mathrm e}/{\mathrm{m_{e}c}^2}$, and the 
energy-dependent optical depth to Compton scattering, $\tau_{\mathrm e}$. 
The coronal spectrum 
can be well approximated by a power-law for a  region of parameter-space 
up to a cut-off energy $E_{\mathrm{c}} \sim {\mathrm k}T_{\mathrm e}$.  Ignoring geometric details, 
measurement of the continuum slope and the cut-off energy could, in principle,
constrain the 
physical parameters of the corona, and combining this with the 
observed luminosity one could estimate the luminosity of the seed photons and 
learn something about the accretion flow. Unfortunately, in practice the 
primary continuum is difficult to isolate in available observational data. 
In contrast to the spectra of accretion disks in AGN 
that peak in the UV band,  those in Galactic Black Hole binary systems 
have a higher  peak temperature, such that the  
Wien tail of the disk emission imposes a signature in the X-ray band that 
can be difficult to disentangle from the other spectral components. 

The AGN corona is thought to exist on size scales less than a few tens of gravitational radii 
from the black hole (section~\ref{continuumsize}), 
so the X-ray continuum could experience reprocessing in gas extending from 
very close to the black hole outwards. 
X-ray continuum photons produced in the corona 
would emerge quasi-isotropically (depending on the corona optical depth and shape), 
some of these escape along sight lines to the observer and 
could pick up an imprint of material along the line-of-sight, some X-ray photons may
illuminate the accretion disk and reflect from the surface of the disk before 
reaching the observer (Figure~\ref{fig:schematic}). 
The observed spectrum is then the sum of primary continuum and 
reprocessed X-rays. While frustratingly making the true continuum difficult to uncover, these 
imprints of 
reprocessing are themselves valuable diagnostics of accreting systems, and we now discuss the 
details and diagnostic potential of those signatures. 

\subsection{X-ray reprocessing in the Compton-thin regime} 
\label{cthin}
In the $0.1 - 10$\,keV band, energetic X-ray photons can excite or
ionize inner-shell electrons of atoms from C to Ni, with K-shell
processes dominating. Early (proportional counter and other)
observations of X-ray absorption in AGN concentrated on deriving gas
parameters from the broad ``curvature'' (the systematic variation in
apparent power-law slope with photon energy)
produced by superposition of
bound-free edges presumed from neutral gas, whose effects were most
noticeable in the soft X-ray band. 
Initial treatment of ionized gas derived column densities and
ionization-states from fits to K edges of O\,{\sc vii} and O\,{\sc viii}.  
As data quality improved, and more detailed spectral features
became detectable, the employment of photoionization models was
justified, although most models necessarily make simplifying
assumptions of geometry and distribution of matter.  Common features
of many models are the assumption of a slab of gas of constant
density, photoionized by a central continuum source whose spectral
form is usually specified as an input parameter.  The gas is often
assumed to be in thermal equilibrium with the thermal and ionization
structure determined at some finite number of points in radial steps
away from the central continuum.  Photoionization codes that have been
widely used to fit X-ray spectra include 
{\sc ion} \citep{netzer93a},  
{\sc cloudy} \citep{ferland98a}, 
{\sc xstar} \citep{kallman01a} (with {\sc xstar} 21ln onwards including the
important inner shell transitions for Fe, \citealt{kallman04a}),
{\sc spex} \citep{kaastra03a},
{\sc titan} \citep{collin04a} and
{\sc mocassin} \citep{ercolano08a}.

In photoionized gas the ionization state is determined by the form of the 
ionizing spectrum and a quantity that measures the 
ratio of ionizing photon density to proton density known as the 
{\em ionization parameter} \citep[e.g.][]{osterbrock06a}
In this review we use the $\xi$ form of ionization parameter where
$\xi=L/nr^2$ and $L$ is the ionizing luminosity integrated from
1-1000 Ryd, $n$ is the proton density and $r$ the distance of
the material from the central black hole 
\citep{tarter69a}\footnote{
The ionization parameter is also commonly defined as a dimensionless 
quantity $U$ \citep{osterbrock06a}, widely used in UV
astronomy: additional variants on the ionization parameter are $U_x$
and $U_{ox}$ that are distinguished by the range over which the
illuminating ionizing luminosity is integrated 
\citep[0.1-10.0 keV and 0.54-10.0 keV, respectively, see e.g.][]{netzer96a,george98a}. 
The conversion between
the most commonly used ionization parameters is shown by
\citet{george98a} for several assumed ionizing spectra.
For analysis of multi-phase plasmas in pressure equilibrium a
parameter $\Xi$ defined
in terms of pressure rather than density may be preferred
\citep{krolik81a, nayakshin00a}.
}
We adopt units for $\xi$ of erg\,cm\,s$^{-1}$.
A convenient expression of the amount of gas in the line-of-sight is the
equivalent column density of hydrogen atoms, ${\mathrm{N_H}}$ atoms cm$^{-2}$,
with solar abundances of other elements often assumed, although ionic columns, 
i.e. column densities of just one species, are also widely used.
In high-ionization gas, hydrogen does not contribute significantly
to the opacity, which is dominated by photoelectric absorption by heavier
ions.

In addition to bound-free edges, resonance transitions can
produce significant absorption signatures.  The importance of resonance
absorption lines in the X-ray regime was first discussed by
\citet{matt94a}, who noted that if material is in an ionization-state
for which resonant absorption can occur 
(i.e. for K$\alpha$, those having an L shell vacancy, H-like to F-like ions) 
then in the optically-thin regime the  Fe\,K$\alpha$ absorption line 
equivalent width (EW) would be
$\mathrm{EW} \simeq 20 A_{\mathrm{Fe}} (\mathrm{N_H}/10^{22}\mathrm{cm}^{-2})$\,eV,
where $A_{\mathrm{Fe}}$ is the iron abundance relative to the interstellar value, 
indicating that Fe\,K$\alpha$ absorption should be 
highly significant in X-ray spectra of ionized gas 
(see also \citealt{nicastro99a} and \citealt{bianchi05a}). 
Indeed, a few years after the
publication of that paper, with the advent of a new generation of
X-ray grating instruments, resonance absorption lines started to be
detected in X-ray spectra of AGN 
\citep[][Figure~\ref{fig:3783lambda}]{kaastra00a, sako01a, kaspi02a}. 
\citet{krolik95a} discussed how the resonance line features would
contribute to the X-ray spectral shape, such that they would seriously
compromise correct identification of simple single edges fit in CCD
quality spectra and also affect the observed fluxes and
profiles of any emission lines close enough in energy to be
unresolved in X-ray spectra.

Absorbing gas is also expected to produce line emission.
In particular, 
K$\alpha$ lines may be emitted by fluorescence, that is, relaxation
after a K-shell
photoionization, with a high fraction of photoionizations ultimately producing
a K$\alpha$ photon \citep{osterbrock06a}.
The strength of an emitted fluorescence line is governed by the probabilities
of radiative de-excitation compared to auto-ionization (known
as the Auger effect for this case of relaxation from a state with
an inner-shell vacancy) and to
two-photon emission. The probability of producing a
K$\alpha$ photon per ionization 
is known as the (K$\alpha$) fluorescence yield.  Because fluorescence
yield increases strongly with atomic number approximately as $Z^4$, 
and because Fe has
a high abundance, Fe line emission is prominent in X-ray spectra of AGN.

\citet{krolik87a} estimate that the luminosity of the Fe\,K$\alpha$ 
fluorescence emission line is approximately
$$
L_{\mathrm{K}\alpha} \simeq \frac{E_{\mathrm{K}\alpha}}{2+\Gamma}
\langle Y \rangle
\frac{\Delta \Omega}{4 \pi} 
L_0
\langle \tau_0 \rangle
$$
in the optically-thin limit and assuming a power-law illuminating continuum
of photon index $\Gamma$, and 
where $E_{\mathrm{K}\alpha}$ is the energy of the K$\alpha$ emission line, 
$\langle Y \rangle$ is the fluorescence yield averaged over ionization states,
$L_0$ is the continuum luminosity per unit energy interval at the K edge
and $\langle \tau_0 \rangle$ is the photoionization
optical depth at the K edge averaged over the solid angle subtended by the 
absorber at the source.  The above relation is an upper limit on the line strength because
scattering and re-absorption of the fluorescence photons have been neglected.  
A consequence of Fe ions having high fluorescence yield is that we expect
to see a direct correspondence between detection of bound-free
edges (such as the Fe\,K edge) and re-emission of fluorescence lines
(such as Fe\,K$\alpha$).  For low-ionization states of Fe we expect 
approximately one-third of photons absorbed in a K-edge photoionization to be re-emitted as
K$\alpha$ line photons.
For highly-ionized material we expect
$0.5 \la \langle Y \rangle \la 0.7$
\citep{krolik87a}.
Because the bound-free edge is observed
in absorption along the line of sight, but the fluorescence photons
are emitted isotropically, the equivalent width of the emission depends
on the geometry of the absorber and in particular on the 
fraction of the source that is covered by the absorber.  Only in the
case of complete covering of the source by optically-thin gas do we
expect to observe an equivalent width commensurate with the 
fluorescence yield.

For ionized material, the binding energies of 
the inner shells generally increase with ionization 
and so the energies of the K edges and emission and
absorption lines usually increase with ionization (see \citealt{kallman04a}).  
In neutral material the Fe\,K$\alpha$
fluorescence line comprises a doublet at
energies 6.404 and 6.391\,keV, these being
indistinguishable by current X-ray spectrometers. The Fe\,K$\beta$ line
at 7.06\,keV is expected at about $13.5\%$ of the flux of K$\alpha$
\citep[e.g.][]{leahy93a,palmeri03a}.
In ionized gas the ``Fe\,K$\alpha$ line'' becomes a complex of permitted,
intercombination and forbidden transitions (depending on which ion is
considered) with possible contributions also from satellite lines,
that are largely unresolved with current instruments 
(see e.g. \citealt{bautista00a}; \citealt{kallman04a}; \citealt{bianchi05a}),
but which may be partly separable with future instruments (see Section\,\ref{sec:prospects}).
In photoionized gas in particular, Fe\,K$\alpha$ fluorescence emission 
may be dominated by the intercombination and forbidden transitions 
(e.g. for Fe\,{\sc xxv}, see \citealt{bautista00a}; \citealt{bianchi05a} 
and references therein, 
see also \citealt{bautista00a}; \citealt{porquet00a} for discussion of other He-like ions).
The fluorescence yield is also a
function of ionization state; the yield varies with ionization
until the atom is down to H- and He-like states, where
the Auger effect can no longer occur and the yield reaches a maximum \citep{krolik87a}.
\citet{matt97a}, \citet{bianchi02a} and \citet{bianchi05a}
give examples of the predicted line strengths and dependences on atomic number,
column density, geometry, elemental abundance and ionization-state 
of the line-emission region. 

Returning to consideration of resonance absorption lines, emission
of K$\alpha$ photons following a K$\alpha$ absorption 
(the process of resonant scattering) can also occur, causing the absorption
lines to be partially filled-in.
Neglecting, for the moment, the destruction or absorption of line photons
that can occur during successive resonant scatterings 
(see Section\,\ref{sec:linestrength})
and assuming the optically-thin limit,
the strength $F_{em}$ of resonance line emission accompanying the
line absorption is expected in Li-like to F-like ions to be approximately 
$F_{em} \simeq F_{abs} Y' \Delta\Omega/4\pi$, where
$F_{abs}$ is the absorbed flux, $Y'$ is the fluorescence yield for the
ionization state below that of the ion in question
and $\Delta\Omega$ the solid angle subtended by the
absorber at the illuminating source \citep{matt94a}.
The fluorescence yield for the previous ionization state, $Y'$, is
close to, but not exactly the same as, the yield of K$\alpha$ photons
from the resonant transition.  For H-like and He-like ions $Y'$ should be
replaced by a value of nearly unity, as K$\alpha$ transitions in these
ions are almost pure scattering events.  The geometry of the
gas is a crucial factor in determining the degree to which the
absorption line is filled-in by re-emission and for a full sphere of
highly-ionized optically-thin 
gas the emitted flux is typically about one-half the absorbed flux
(but see below for further discussion of the effects of geometry on 
resonantly-scattered lines). 
Clearly the relative dominance of various emission and absorption
features depends strongly on the gas geometry, and so conversely, the
measured strengths of these could in principle be a powerful constraint on the gas
distribution, if components can be adequately separated in
observational data.

\subsection{Resonant line scattering and the effects of geometry}
\label{sec:linestrength}

In ionized material,
for radiative transitions whose lower level is significantly populated,
the photon mean free path may be much less than
for continuum photons, and under these conditions 
line photons can resonantly scatter (e.g. the Fe\,K$\alpha$ resonance line
in the case where Fe is ionized at least to F-like states).
Losses of line emission may occur for lines such
as Fe\,K$\alpha$ during resonant scattering. First, on each scattering there
is some probability of photoelectric absorption, so
on repeated resonant scattering events the line becomes increasingly attenuated.
Second, for Li-like to F-like ions, at each resonant absorption there is a probability
of auto-ionization rather than remission of a K$\alpha$ photon - a process often
referred to as ``resonant Auger destruction''.
Detailed calculations of population levels
are required to correctly assess the extent both of resonant scattering and
 of resonant Auger destruction
\citep[e.g.][]{liedahl05a} but the effect can be significant for ionized
absorbing zones.  The extent of resonant scattering is also affected by
the velocity structure in the gas: turbulent velocity structure lowers the 
optical depth in the line core but increases the range of photon energies over which
resonant scattering can take place \citep{nicastro99a}; macroscopic velocity
structure can suppress resonant scattering, which ceases when
line photons are Doppler-shifted away from the wings of the absorption profile.

The geometry of the absorbing material
can play an even more important role for resonantly-scattered lines than it does for 
non-resonantly scattered lines.  Not only is the covering fraction important but so also
is the shape of the absorbing zone and orientation with respect to both 
illuminating source and observer.
Consider radiation passing through an anisotropic
absorber.  When resonantly scattering, line photons
random walk between successive absorption and emission events inside
the absorber, and are last scattered from about an optical depth 
of unity inside its volume.  
If the absorber is oriented so that its surface is preferentially
oriented away from the observer, there is a greater line intensity
away from the observer than towards the observer.  The effect has been
simulated for resonant scattering in highly ionized gas by \citet{matt97a}.
In addition, if the K-edge absorption optical depth is sufficiently high, most
fluorescence photons are emitted from the surface nearest the illuminating
source, again leading to anisotropy of the line emission, with
the observed luminosity being significantly suppressed for lines-of-sight
that pass through the absorber \citep{ferland92a}.

Given that there is no spatial information available
for the innermost X-ray-emitting regions, the geometry of the regions
is unknown, making predictions of the observed strengths of resonantly-scattered lines such as
ionized Fe\,K$\alpha$ difficult and model-dependent.

\subsection{X-ray reprocessing in the Compton-thick regime}
In gas with a high column density, Compton scattering has a significant
effect.
For photons in the 2--10\,keV band, where most X-ray data have been
accumulated to date, h$\nu << \mathrm{m_e c}^2$ and the Compton
scattering cross-section for free electrons is approximately the
energy-independent Thomson cross-section $\sigma_{\mathrm T}$.  For
solar-abundance gas in which H and He are fully ionized, the dominant
scattering is by free electrons and the continuum optical depth is
$\tau \simeq \mathrm{N_e} \sigma_{\mathrm T}$, where $\mathrm{N_e}$ is
the free electron column density (heavier ions also have significant
energy-dependent individual scattering cross-sections but their net contribution is
small).  Thus the scattering optical depth has a value unity
approximately at ${\mathrm{N_H}} \simeq 1/1.2\sigma_{\mathrm T} \simeq 1.25 \times
10^{24} {\mathrm{cm}^{-2}}$. For solar abundance material, the
cross-sections for Compton scattering and photoelectric absorption,
$\sigma_{\mathrm{pe}}$, are comparable at 10\,keV and this represents
an energy that has been considered as the low threshold for studying
the ``Compton reflected'' components of AGN.  For column density
up to about $10^{25} {\mathrm{cm}}^{-2}$, some transmitted
radiation is still visible above 10\,keV \citep{matt99a}. Note that
in this regime, computation of the expected transmitted spectrum should
take account of Compton scattering, which is not included in 
one-dimensional radiative transfer codes such as {\sc xstar}.
For higher
column densities still, the material allows essentially no direct
transmission of photons through the structure.  Compton scattering
around the edges of absorbing zones may still allow some fraction of
X-ray radiation to reach the observer (rather like the
``silver-lining'' effect around cloud edges).

Despite the presence of an intense field of radiation, relatively cold
gas ($T \leq 10^6$K) can exist in the nuclear regions if the
material is sufficiently dense \citep{guilbert88a,ferland88a}. It is
thought that the inner radius of the accretion disk may extend very
close to the black hole, and that this disk could provide a
Compton-thick structure that plays an important role in reprocessing
nuclear X-rays.
Photons incident on the disk surface penetrate to a mean optical
depth of unity into the disk.  Continuum X-ray photons undergo either
Compton scattering or photoelectric absorption in this surface layer.  
Compton scattering can redirect photons
back out of the slab; in such an encounter, the typical Compton recoil energy loss
$\Delta E$ by a photon of energy $E$ 
is given by $\Delta E/E \simeq 2 E/{\mathrm{m_e c}}^2$
which exceeds 10 percent at $E \ga 30$\,keV.
Incident photons are downscattered in energy and a ``hump''
appears in the reflected spectrum through a combination of scattering and absorption effects. 
If we view the disk from the same
side as the illumination there are not many scatterings per observed photon
and for photon index $\Gamma=2$ the hump is at most a factor two enhanced 
\citep{magdziarz95a} compared with a calculation in which Compton
losses are neglected.
Below 30\,keV the photoelectric opacity of the material increases with decreasing 
energy, further enhancing the appearance of the reflection hump \citep{matt91a}.
A photoelectric absorption event results in either
fluorescence line emission or ejection of an Auger electron.  Thus the
fate of incident photons is either to be destroyed by the Auger effect, scattered
out of the slab or re-emitted from the slab as a fluorescence line (with some
small contribution from two-photon emission).
The net effect of these processes
gives the so-called Compton reflection spectrum.

The shape of the reflection spectrum has been determined from theory
\citep{guilbert88a,ferland88a} and calculated in Monte Carlo
simulations \citep{george91a,reynolds97b,matt91a}.  The combination of
absorption and scattering reduces the intensity of the reflected
spectrum compared to that incident on the disk.  
To first order, valid for energies $\la 10$\,keV, the reflected spectrum is depressed 
by an energy-dependent factor about 
$\sigma_{\mathrm T}/(\sigma_{\mathrm T}+\sigma_{\mathrm{pe}})$ 
relative to the illuminating continuum and is a function of the ionization
of the surface layers in the reflector.  At energies $\ga 10$\,keV the
Compton ``hump'' discussed above enhances the reflected spectrum somewhat.
The observed strength of the reflected spectrum  
relative to the illuminating continuum is
a strong function of illuminating angle and viewing angle
\citep{ghisellini94a,george91a,matt97a}, being highest for
viewing normal to the surface, and,
for an infinite disk illuminated by a point source,
varying by a factor about $10$ as 
viewing angle $\theta$ varies over the range $0.05 < \cos\theta < 0.95$
(\citealt{magdziarz95a}: see also \citealt{nayakshin00a}). 
In model fits to data
the relative reflection strength at high energy (where 
the opacity is low) is often parameterized by a parameter $R$ defined
as the ratio of the reflected intensity relative to that expected from
a uniform infinite disk with time-steady illumination from above and viewed 
normal to the surface: a surprising feature of some model fits to
data is that they yield  values $R \ga 3$ 
\citep[e.g.][]{miniutti07a}, see Section \ref{sec:hardspectra}.  

X-ray missions to date have offered limited data above 10\,keV, making
it difficult to access this important part of the spectrum (without
going to high redshift, where a dearth of photons limits progress).
Fortunately however, there are other detectable signatures of
reflection predicted below 10\,keV.  As noted above, some absorbed
photons are re-emitted as fluorescence lines and the emission part of
the spectrum is dominated by the K$\alpha$ lines of the most
abundant metals.  Fe K$\alpha$ is the strongest of these fluorescence
lines and was predicted to have a detectable contribution from cold dense 
material close to the black hole whether in the form of clouds  \citep{guilbert88a,nandra94b} 
or an accretion disk \citep{lightman88a}. 
At softer energies, emission is expected from
other fluorescence lines, radiative recombination continuum and
bremsstrahlung from the reflector surface layers leading to a 
rich spectrum that is highly dependent on the ionization
parameter \citep{ross99a, ross05a}.

As noted above, the photons that escape from the disk are those 
that reprocess within an average of one optical depth of the surface, 
and thus it is the properties of the skin of the disk that 
determine the detailed shape of the reflected spectrum, i.e. the
reflection spectrum can have the signature of ionized material without
inferring any inconsistency with the conditions required for the disk 
to exist at that location.  The strengths of the spectral features in
the case of reflection also depend on the factors that are important for
Compton-thin gas, i.e., 
elemental abundance, ionization-state, incident spectrum
and geometry (in this case, the shape and orientation of the disk).
Destruction of line photons
during resonant scattering (Section\,\ref{sec:linestrength}) also
occurs: for the Fe\,K$\alpha$ emission this is important for the
range 100\,erg\,cm\,s$^{-1} \la \xi \la 500$\,erg\,cm\,s$^{-1}$, where
Fe is dominated by ions in the range Fe\,{\sc xvii-xxiii} \citep{kallman04a}.
As $\xi$ increases above this level photoelectric absorption
and the Auger destruction diminish, and although
resonant scattering is still in effect, the line photons can
eventually escape the disk and the observed line strength increases, up
to a point where Fe is completely ionized and both line and edge
disappear from the reflected spectrum (see \citealt{ross99a, ross05a}).

The X-ray spectra of AGN are likely complex, with contributions from a
primary continuum that may have a high-energy cut-off (thought to be
in the few hundred keV regime, \citealt{dadina08a}), absorption from several layers
of gas with different ionization and likely differing covering
fraction, and a contribution from reflection off Compton-thick
gas. On this is superimposed the re-emission associated with each
absorption and reflection process, and those components are
subject to distortion by various kinematic and relativistic effects. 
We now consider the distortions important to the X-ray regime.

\subsection{Thermal, kinematic and relativistic modifications of AGN spectra}

The modification of the reprocessed
spectrum by a variety of kinematic and relativistic effects
complicates the effort to identify the physical state and geometry of
reprocessing material.
Consider first the effects that are expected for a cloud of gas
experiencing no external effects.  For the Fe K$\alpha$ fluorescence
line the natural line-width is about $3.5$\,eV, negligible given the
spectral resolution available and given many other factors that
broaden the line far more.  Thermal broadening in gas of temperature
$T$ produces a Doppler spread in photon energy $\Delta E/E \sim
\sqrt{2 {\mathrm k}T/{\mathrm{m_{p}c}}^2}$. For $T\sim
10^6-10^7$\,K, likely applicable to the X-ray gas, thermal widths are
also negligible compared with instrumental resolution, although thermal
broadening does affect the extent of resonant scattering of lines.

Compton scattering can also have a detectable effect on line profiles. At
Fe\,K$\alpha$ a single Compton scattering event can change the photon energy
by an average of about $80$\,eV, comparable to the resolution of CCD detectors and resolvable 
by grating instruments, and leading to the possible development of a ``Compton
shoulder'' to energies below the line core.  

Other observational distortions depend on the gas geometry,
radial location from the central black hole and our viewing
line-of-sight.  Spectral features imprinted from gas with some
velocity are shifted in energy; in the non-relativistic regime this
is a simple Doppler shift.  The case of the
symmetric sphere of outflowing gas is interesting and well studied
within stellar astrophysics; the so-called P\,Cygni profile is produced in
such a case, where the absorption component of a line is blue-shifted
relative to the mean emission component, producing a characteristic spectral
signature. 
In all cases except that of the symmetric sphere of gas, the observer's
orientation to the material is clearly important as that determines
the velocity component along the sight-line and hence the degree of
Doppler shift experienced by each line component. 
For Compton-thin gas, absorption features can be seen from material
in the line-of-sight to the continuum, and in this case broadening can 
occur  where there is a velocity gradient along the line-of-sight.
Line emission is similarly broadened both by bulk flow velocity
gradients and by orbital motion.
In some regimes of inclination angle and emitting radii, 
gas in a rotating annulus may produce emission lines with the classic
``double-horned'' profile, as the observer is viewing the sum of the
approaching and receding sides of the disk.

As gas velocities reach the relativistic regime then spectral
signatures suffer further distortion, as relativistic aberration
boosts the blue peak of spectral lines and suppresses the
red.  For emitting material very close to the black hole, gravitational
redshift of photons becomes significant, redistributing line
photons to lower energies with a shift that depends on the radial
location of the gas; in addition to gravitational redshift the
Doppler effect from orbital motion provides an energy shift of comparable
magnitude; and light bending around the black hole leads to potentially
large distortions in the viewed emission region.  In the case of emission
from a disk, the latter effect is most significant for disks viewed at
large inclination angles to the normal.
\citet{fabian89a}, \citet{laor91a}, \citet{karas95a} and  \citet{reynolds03a},
among others, plot the predicted spectra from an accretion disk, showing the 
expected line profile as a function of disk orbital parameters.
As the radial location of the gas is a critical parameter
for determining the degree of general relativistic effects, this, and
other key parameters can, in principle, be derived from spectral line
shapes.
In the Schwarzschild metric, the general relativistic terms 
are factors $(1 - 2{\mathrm{r_g}}/r)^{-1/2}$ which become significant
for radius coordinate $r \la 20$\,\rg.  Thus 
general relativistic distortions are, in principle, measurable in
current X-ray spectra for gas within such radii, although, as we
discuss below, the ambiguity of spectral signatures limits progress in
this regard.

\section{Basic Models of AGN spectra}\label{sec:basic}
\begin{figure}
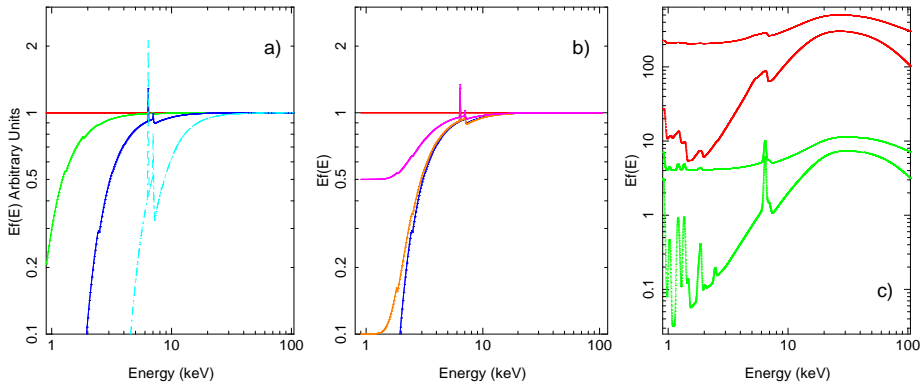

\begin{center} 
\hbox{
\includegraphics[width=40mm, height=50mm]{fig2a.eps}
\includegraphics[width=40mm, height=50mm]{fig2b.eps}
\includegraphics[width=40mm, height=50mm]{fig2c.eps}
}
\end{center}
\caption{\footnotesize
{Model lines for a) a simple power-law continuum with photon index $\Gamma=2$, no high-energy cut-off 
and no absorption  (red),  modified by 
absorption by a full shell of neutral gas with column density 
${\mathrm{N_H} = 5 \times 10^{21} \mathrm{cm}^{-2}}$ (green), 
${\mathrm{N_H} = 5 \times 10^{22} \mathrm{cm}^{-2}}$ (dark blue), 
${\mathrm{N_H} = 5 \times 10^{23} \mathrm{cm}^{-2}}$ (light blue dot-dash line); 
b)  the same power law  (red) absorbed by a full shell of gas with 
${\mathrm{N_H} = 5 \times 10^{23} \mathrm{cm}^{-2}}$ (dark blue), 
90\% covered by the gas (orange) and 50\% covered by the 
gas (magenta); c)  the spectrum reflected by Compton-thick
gas calculated from the {\sc reflionx} model \citep{ross05a}.
We plot reflection alone (lower
green line) and summed with the illuminating powerlaw (upper green line).
The same reflection spectrum is shown blurred by the effect of
orbiting of the material near the black hole (lower red line) and 
the sum of the blurred reflected
spectrum and the powerlaw is also shown (upper red line). 
The two cases are each shown with an arbitrary normalization to allow clarity of
display.
See text for full details of the calculations.
}}
\label{fig:cartoons}
\end{figure}

The above basic ingredients may be used to construct model spectra
with a small number of parameters that may be used to fit to observations
(see Sections\,\ref{sec:early} \& \ref{sec:recent}).  Fig.\,\ref{fig:cartoons}
illustrates the model components that are commonly used, all shown here for
the same simple input continuum, a power-law with photon index $\Gamma=2$.
The first panel (a) shows the effect of a shell of absorbing gas in front of the
primary source.  In this case the gas is assumed to be neutral and hence has
very high optical depth in the soft band.   
As discussed above, models to generate the effects of ionized absorption
are also frequently used (e.g. {\sc xstar}, 
\citealt{kallman01a, kallman04a}\footnote{http://heasarc.nasa.gov/lheasoft/xstar/xstar.html}).
Curves show the transmitted spectrum for a full shell of gas of varying
column density and the Fe K$\alpha$ and K$\beta$ line emission 
from the gas is also included based on the calculations of \citet{leahy93a},
assuming the K$\beta$ line flux to be 13.5\% of the K$\alpha$ line.   
The second panel (b) shows the effect of allowing only part of an
extended source to be covered by the absorbing gas, showing covering fractions
of 50, 90 and 100\,percent.  The dominant effect is to allow leakage of flux  
in the soft band, and to change the shape of the transmitted spectrum 
at intermediate energies. Emission lines are included as for (a). 
The third panel (c) shows the spectrum expected
from reflection from optically-thick gas, here generated using the {\sc reflionx} model 
\citep{ross05a} assuming $\xi=10$\,erg\,cm\,s$^{-1}$ and solar abundances.
Also shown is the sum of the reflected spectrum 
and the illuminating powerlaw. For comparison, these are 
overlaid with the spectrum expected after
general relativistic blurring appropriate to reflection within 3--20\,\rg\ of the black hole, 
with a radial profile of reflected emissivity
$I(r) \propto r^{-3}$ and assuming in the blurring calculation an inclination angle of 30$^{\circ}$.
This calculation was made using the {\sc kdblur} convolving function, based on the model
of \citet{laor91a}, in the {\sc xspec} package \citep{arnaud96a}.

These models are overly simplistic representations of the likely complex
inner regions of AGN.  Any absorbing material is unlikely to be in the form of a uniform shell
and probably comprises multi-phase gas over a wide range of ionization, density and
radii, and having spatial structure which, if comparable to the scale of the
primary emission region, causes a mixture of spectral components to be received by
the observer.  Any absorbing gas of high optical depth also produces
scattered components of radiation that should be taken into account.
Similarly, although we expect there to be scattered/reflected radiation from 
optically-thick components in the inner regions,
this is likely complex, again covering a wide range of radii, ionization states and
spatial structures.  \citet{nayakshin00a} show how an accretion disk atmosphere in pressure
equilibrium may produce substantially different reflection spectra from the constant-density
calculation used for {\sc reflionx} \citep{ross05a} for some range of disk states.
Also, although the approach noted above to construct the reflection models is 
one that is commonly used, the 
one-dimensional calculation used for {\sc reflionx} does not allow for
the expected variation in line emissivity with both illumination incidence angle
and viewing angle \citep[e.g.][]{nayakshin00a} and thus the calculation for inclined disks 
is not self-consistent.  For inclination angles $\la 30^{\circ}$ the variation in
emissivity is not thought to be large \citep{magdziarz95a} but would make a
significant difference at inclinations $\ga 60^{\circ}$.  Near the black hole the combined
effects of relativistic aberration of angles caused by the disk motion and
the bending of light around the black hole may make such angle-dependent
emissivity more important.
Furthermore, although we adopted a reflected emissivity profile varying as $r^{-3}$,
we do not in fact expect the reflection  
to vary with the same dependence as the black-body emissivity 
of the disk \citep{shakura73a}, as
the reflected emissivity depends on the radial dependence of the illuminating
continuum, e.g. radiation from a corona, viewed at the disk's surface,  
as well as on the disk's structure and ionization.  
A large, extended corona would not be expected to result
in such a steep reflection profile.

With the datasets currently available it is not possible to
build models that can encapsulate such complexity in any unambiguous way, so the best
that can be done for now is to try to model the generic features of the X-ray spectra
to distinguish the most important spectral components.  This has important 
consequences for the conclusions that may be drawn from model-fitting: if we wish to
decide between two competing models of a spectrum, in reality either of the models
being tested are likely gross simplifications of the true physical situation.  A simple
comparison of overall goodness-of-fit, e.g. using a standard $\chi^2$ test, could be
misleading.  A better approach under these conditions is to attempt to identify key
features, such as absorption lines or edges or emission lines, which might have
the power of discriminating between models, and to test the data for the presence of
those signatures.  As the quality of data
improves with each successive X-ray observatory, our ability to distinguish between 
such signatures increases.  But even with current generations of instruments, CCD detectors with 
high effective area lack the energy resolution needed to clearly detect narrow absorption
and emission features, whereas high resolution grating instruments lack sensitivity.
This results in an ongoing ambiguity in the conclusions that may be drawn from modeling
of X-ray spectra, but this ambiguity should be addressed in the next few years with
a new generation of high-throughput, high-resolution and broad-band instruments
(Section\,\ref{sec:prospects}).
In the following two sections we take a historical approach to
describe how our knowledge of the inner regions of AGN has developed to date, and to summarize
some of the current ambiguities in the interpretation of AGN X-ray spectra.

\section{Early X-ray observations of AGN}
\label{sec:early}

In this section we summarize some of the strengths and notable properties of the 
X-ray missions that have had the most major impact on AGN research.  

Observations using balloon and rocket-borne detectors flown during the
1960s detected X-rays from extra-galactic sources such as 3C 273 and
Centaurus A \citep{bowyer70a}.  Following these successes, several
dedicated X-ray astronomy satellites were approved that operated
during the 1970s. One of these, the {\it Uhuru} satellite, confirmed
the X-ray detection of several sources identified with AGN \citep{giacconi74a}.  
{\it Ariel-V} data
\citep[e.g.][]{elvis78a} later showed that not only was X-ray emission
a common property of AGN but that the X-ray flux
varied by significant factors down to timescales shorter than one day in
some sources \citep{marshall81a}.  Of the many instruments on {\it
 OSO-7} and {\it HEAO-1}, the proportional counters were the most
useful for AGN studies. These, along with the {\it Ariel V} data,
established that in the $2-20$\,keV regime AGN spectra could be
parameterized by power-law continua with a mean photon index for the
sample $\Gamma \sim 1.7$ (e.g. \citealt{tucker73a},
\citealt{mushotzky76a}, \citealt{mushotzky84a}).  Column densities
were estimated assuming the absorbing gas to be a shell of neutral
material covering the illuminating source. Compilations of spectral
fits suggested column densities ${\mathrm{N_H}} \sim
10^{21}$\,cm$^{-2}$ of gas existed around active nuclei, that this gas
may be located in the BLR and that obscuration is more likely for low
luminosity objects \citep[e.g.][]{lawrence82a}.

The second {\it HEAO} satellite, launched in 1978, was named the {\it Einstein} observatory. 
{\it Einstein} provided  focusing X-ray optics and high sensitivity whose combined properties 
increased the number of X-ray detections of AGN by an order of magnitude. The  
focal plane detectors included proportional counters and a solid state spectrometer; 
observations using these instruments revealed soft-band (here used to mean 0.1--2\,keV) 
complexity \citep[e.g.][]{wilkes87a} 
that led to some suggestions that the 
gas covers only  a fraction of the line-of-sight of AGN, comprising a so-called `partial-covering absorber'  
(\citealt{reichert85a,holt80a}; see Section~\ref{sec:basic}). 
{\it Einstein}  observations of variations in  absorption for the quasar 
MR 2251-178 led \citet{halpern84a} to the first suggested detection of a 
partially ionized absorber in AGN, dubbed the `warm absorber', 
responding to variations in the continuum source.  

The European X-ray Observing Satellite, {\it EXOSAT}, had a 3 year mission (1983- 1986) and provided 
 a variety of instruments of which the channel multiplier array  (CMA) and the 
 medium energy (ME) proportional counter were most useful for studies of AGN. 
The satellite had focusing optics in two Wolter type 1  Low Energy Telescopes each with a CMA in the focal plane 
that yielded data in the 0.05--2\,keV band.  
As for previous proportional counters, 
the ME  provided only  modest spectral energy resolution 
($\Delta E_{\mathrm{FWHM}}/E \simeq 0.21 (E/6.0{\mathrm{keV}})^{-0.5}$ for the argon chambers) 
and useful data over 1--10\,keV for weak sources such as AGN.  
{\it EXOSAT} also carried  grating detectors (the Transmission Grating Spectrometers) but the 
sensitivity of the detector  combination was not sufficiently high for any useful AGN observations to be 
carried out.
Some notable contributions of {\it EXOSAT} to the study of AGN  included detailed study of rapid variability 
in AGN \citep{green93a}.  
Timing studies of AGN were aided by the eccentric orbit that  allowed 
continuous observations  for up to 76 hours of the 90 hour orbit.  
The broad bandpass of  {\it EXOSAT}
was also key to the detection of an excess of soft-band  
X-ray flux (the so-called `soft excess') seen in about $50\%$ of AGN  
compared to an extrapolation of model fits to data above a few keV \citep{turner89a}. 
Some authors discussed how the reduced soft-band opacity of partially ionized 
C, N, O, Ne and other ions might explain the soft spectra and variability  observed 
\citep[e.g.][]{fiore90a,warwick88a,yaqoob89a,pan90a} 
while others argued for an interpretation of the soft excess as 
emission closely related to the inner accretion disk \citep[e.g.][]{arnaud85a, kaastra89a,piro88a,turner88a}. 

In addition to the soft-band spectral complexity, some 
detections of Fe emission had been made in the brightest and most heavily absorbed Seyfert galaxies from 
{\it OSO-8} data \citep{mushotzky78a}, the Japanese {\it Tenma} satellite \citep{miyoshi86a} and
some {\it EXOSAT} data \citep[e.g.][]{ghosh92a,leighly89a}. Fe emission lines found in  {\it Einstein} 
spectra \citep{holt80a} were suggested to originate in the BLR, from the same gas 
observed to be providing partial-covering absorption of the X-ray source. 
 
{\it Ginga} (launched by Japan in 1987) provided a 
large area proportional counter yielding good AGN spectra over the range
$2-30$\,keV, whose data established   
Fe\,K$\alpha$ fluorescence emission to be a common property of local Seyfert galaxies 
\citep{pounds89a,pounds90a,matsuoka90a,nandra94a}. Seyfert galaxies were detected 
at energies as high as $20$\,keV, and the 
detection of spectral hardening above 10\,keV \citep{nandra89a,piro90a,matsuoka90a} 
and deep Fe\,K edge were consistent with an origin in neutral gas and
supported the picture of a strong contribution from a reprocessed X-ray 
component. Various origins were discussed for the Fe\,K$\alpha$ line, 
including the putative molecular torus, 
the BLR clouds and the accretion disk. In addition to the earlier suggestions of an origin 
in BLR clouds, a contribution was suggested from reflection off Compton-thick material 
out of the line-of-sight. 
The predicted Compton reflection component seemed consistent with some observed attributes of Seyfert galaxies 
\citep[e.g.][]{guilbert88a,lightman88a,george91a,matt91a}. Several physical  
origins of the Compton-thick material were suggested, including the accretion disk \citep{pounds90a} and the 
molecular torus \citep{krolik87a}. However, the predicted spectra of AGN with 
partial-covering absorbers also 
provided a good explanation of the data \citep{matsuoka90a,piro90a,piro92a}. \citet{celotti92}
expanded on the \citet{guilbert88a} treatment of clouds by consideration of magnetic confinement, finding clouds 
in the density regime $10^{17}-10^{19}{\rm cm^{-3}}$ could survive the intense irradiation likely to be experienced near 
the black hole.  
The hard-band properties of AGN were confirmed by  missions such as 
the {\it Rossi X-ray Timing Explorer} (a proportional counter mission) 
and the Italian mission {\it BeppoSAX} (featuring proportional counters and a collimated Phoswich detector 
system giving a total overall bandpass 0.1--300\,keV), although the ambiguity between Compton-thick reflection 
and complex absorption remained unresolved \citep[e.g.][]{matsuoka91a}. 
{\it Ginga} spectra also showed absorption edges from ionized species of Fe ({\sc xxiv-xxvi}) in  
about $50\%$ of Seyfert spectra \citep{nandra94a}, indicating that a significant column density  
(${\mathrm{N_H}} \sim  10^{23}$\,cm$^{-2}$) of very highly ionized material exists in these nuclei (a discovery 
that was to be confirmed by later grating observations).  

The German-led {\it Roentgensatellit}, {\it ROSAT}, was launched in 1990, 
carrying a telescope of high spatial resolution 
with focal plane instruments covering the soft X-ray band. Individual 
absorption features were found in the  Position-Sensitive Proportional
Counter (PSPC) spectra covering the 0.1--2\,keV  regime, including detection 
of an absorption feature at 0.8\,keV  in MCG--6-30-15 \citep{nandra92a}.  Several other 
bright Seyfert\,I galaxies had sufficiently strong individual features that 
distinct zones  of ionized gas could be isolated 
\citep[e.g.][]{nandra93a,turner93a,turner93b,pounds94a,ceballos96a}.
Most individual features detected using {\it ROSAT} were interpreted as 
a blend of  O\,{\sc vii}\,739.3\,eV and O\,{\sc viii}\,871.4\,eV bound-free edges. 
The soft absorption feature in MCG--6-30-15 was sufficiently well defined that it could be monitored and 
showed variations that indicated 
changes in absorbing gas opacity on timescales from about a day to weeks  
\citep{fabian94a,reynolds95a,otani96a}. 
As the mission progressed, further AGN  observations confirmed the importance of ionized gas in Seyfert galaxies 
\citep[e.g.][]{mihara94a,ptak94a,weaver94a,yaqoob94a,guainazzi94a}. 
{\it ROSAT} observations of  narrow-line Seyfert\,1 galaxies
showed some cases of extremely large amplitude variability, apparently 
exceeding what might be reasonably expected from variable continuum processes \citep{boller97a,brandt99a} 
and arguably favoring X-ray absorption changes as the origin of the observed flux range. 
Motivated by such striking evidence for the importance of absorption variability in AGN, \citet{abrassart00a} 
investigated how X-ray variability might be shaped by clouds partially covering the continuum, finding that 
random rearrangements of the cloud distribution could produce large amplitude variations on timescales of 
$10^2-10^6$s.

The launch of the Japanese 
{\it Advanced Satellite for Cosmology and Astrophysics}, {\it ASCA}, \citep{tanaka94a} in 1993 
offered a significant  improvement in spectral resolution with 
the first flight of CCDs on an X-ray observatory. 
The CCDs formed the Solid-State Imaging 
Spectrometers (SIS) \citep{burke93a} and  offered the valuable  combination of 
improved  spectral resolution ($\Delta E/E \simeq 2\%$ at launch)  
and improved detector sensitivity with low detector background.  While imaging optics had been in use for several years, 
{\it ASCA} was the first observatory allowing imaging in the hard X-ray regime. 
{\it ASCA} also carried a pair of gas imaging spectrometers (GIS) of lower spectral resolution.  
The SIS and GIS allowed determination of 
the X-ray background in an offset region of the detector (clear of target source photons), 
simultaneous  to the accumulation of source counts.   Prior to {\it ASCA}, hard-band detectors had relied 
on the use of non-simultaneous background determination, measurements made in a different detector part or 
the use of background models. The SIS CCDs allowed 
separation of some strong lines and edges that had previously been unresolved in 
proportional counter spectra, with the potential for constraining the shapes of very broad features.  
Of particular interest was the possibility of measuring strong distortions (blurring) of the Fe\,K 
emission line contributions produced within about 20\,\rg of the black hole,
provided the line was strong and not confused with other spectral signatures. 
Indeed a strong spectral curvature around 6\,keV was observed in many Seyfert galaxies and an interpretation made 
as an Fe K emission line, heavily broadened and redshifted by relativistic effects close to the event horizon
\citep{tanaka95a,nandra97a, fabian00a}. If one could be sure of the identification of this spectral component then 
a wealth of diagnostics could be obtained regarding the black hole and accretion disk \citep{fabian00a,reynolds03a}. 
The ambiguity of interpretation of X-ray spectra 
continued to be an issue however; blurred reflection, unblurred reflection and models of 
complex absorption all fit the mean spectra of AGN at a statistically similar level 
\citep[e.g.][]{gondoin01a,reeves04a,turner05a}. 

{\it ASCA} performed some long observations that revealed spectral variability in both the overall observed 
shape and in individual features. For example,
the absorbing gas in MCG--6-30-15 was confirmed 
 to be variable by tracing changes in the O edges first observed using {\it ROSAT}. 
{\it ASCA} data showed 
the depth of the O\,{\sc viii} edge to be anti-correlated with continuum flux, 
while  the O\,{\sc vii} edge appeared unresponsive to variations in continuum flux. \citet{otani96a} 
suggested that the absorber must be composed of 
at least two distinct gas zones, one of which has a significant recombination time  
(see also \citealt{orr97a}). 
Towards the end of the 1990s  a sufficient number of objects had 
been observed by {\it ASCA} and {\it BeppoSAX} that sample studies could start to address 
the overall absorption properties of AGN. \citet{george98a} and \citet{reynolds97a} agreed that 
about $60\%$ of Seyfert\,I galaxies exhibited soft X-ray spectral features consistent with 
ionized gas having ${\mathrm{N_H}} \sim 10^{21} - –10^{23}$\,cm$^{-2}$ 
and $\xi \sim 10 - 50$\,erg\,cm\,s$^{-1}$. \citet{nandra97a} discussed the hard X-ray properties of 
Seyfert 1 galaxies concentrating on the 
apparent profile of the Fe K$\alpha$ line and discussing that 
in the context of blurred reflection models. \citet{turner97a} 
discussed the X-ray parameters of Seyfert 2 galaxies, finding some surprising similarities 
between Seyfert type 1 and 2 AGN with regard to hard X-ray variability and to the structure in the Fe K regime. 

In light of the growing evidence for the importance of ionized gas in AGN systems, some authors revived  
the idea that partial-covering can explain  observed spectral shapes and variability in AGN 
(Section\,\ref{sec:absn}).   
{\it ASCA} obtained data for key sources such as MCG--6-30-15, where \citet{inoue03} concluded  
that a significant part of the broad  
feature observed between 5--7\,keV was the 
result of the superposition of ionized absorbers: this result 
was supported by analysis of the r.m.s. variability of the Fe line band compared with 
the rest of the spectrum \citep{matsumoto03}. 

It is obvious that leaps in understanding have come from each new
development in instrumentation.  The early proportional
counters provided low spectral resolution and often limited spatial
resolution together with a need to take non-simultaneous background
measurements. CCDs allowed some measure of the ratios and variability
of broad and strong spectral features, although the detection of narrow
features such as resonance lines was impossible. The next big steps
came with the advent of grating spectrometers, as we discuss in the
next section.

\section{Current Observatories}
\label{sec:current}
In 1999 two new satellites were launched, the {\it Chandra} X-ray
observatory and {\it XMM-Newton}, offering major steps forward in 
observational capabilities.  With these missions the X-rays incident upon the 
 grating 
instruments were dispersed on to sufficiently sensitive detectors that
grating-resolution AGN spectra could be accumulated with useful
signal, for the first time.

\begin{figure}
\begin{center} 
\hbox{
\includegraphics[width=110mm,height=120mm]{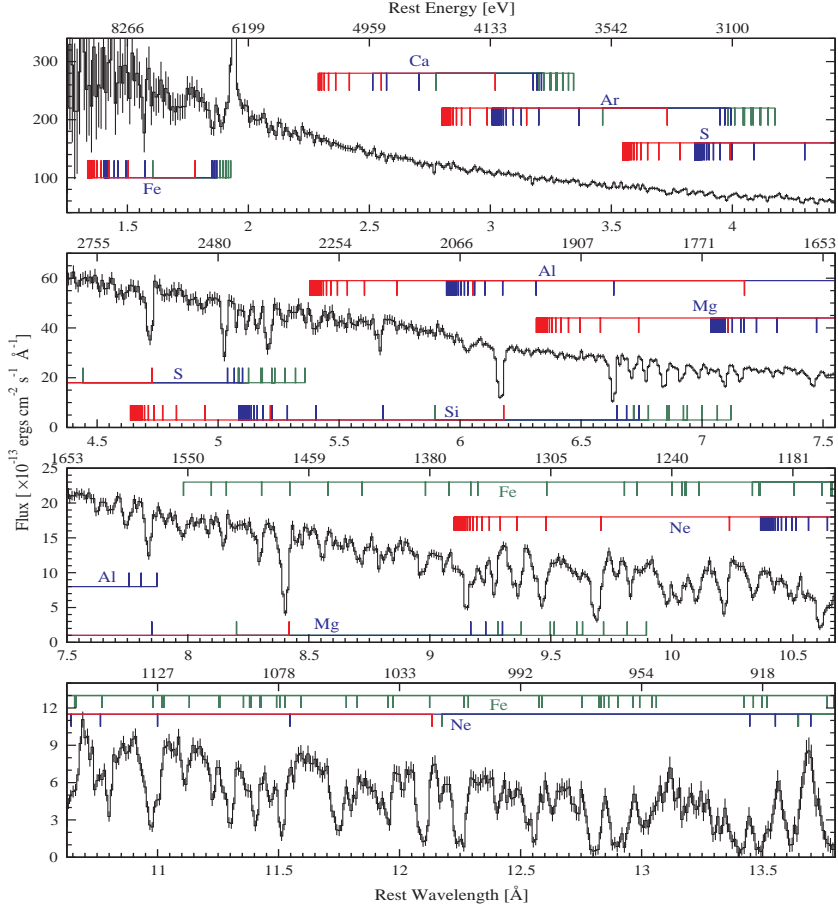}
}
\end{center}
\caption{\footnotesize
{NGC\,3783, summed data from a 900\,ks {\it Chandra} HETG observation. 
The combined MEG and HEG first-order spectrum has been binned to 0.01 \AA. 
Each data point has an error bar representing its 1$\sigma$ uncertainty. 
The H-like and He-like lines of the identified ions are marked in red and 
blue, respectively. Lines from other ions (lower ionization metals and 
Fe\,{\sc xvii}, Fe\,{\sc xxiv}) are marked in green. For each H-like or 
He-like ion the theoretically expected lines are plotted up to the ion's 
edge (not all lines are identified in the data). The ions' lines are 
marked at their expected wavelengths in the systemic rest frame of NGC\,3783, 
and the blueshift of the absorption lines is noticeable. 
{\em From \citet{kaspi02a}: reproduced by permission of the AAS.}
}} 
\label{fig:3783lambda}
\end{figure}

\subsection{Chandra}
The {\it Chandra} X-ray observatory carries  high-resolution optics, providing 
0.5$''$ spatial resolution on-axis (an order of 
magnitude improvement over the best previous missions).
{\it Chandra} provides 
the High Energy  Transmission Grating (HETG) comprising  
two grating assemblies -
the High Energy Grating (HEG) and the Medium Energy Grating
(MEG), calibrated to an absolute wavelength accuracy about $100$ km s$^{-1}$. 
HEG provides spectral resolution FWHM $\Delta \lambda=0.012 $\,\AA\ 
and calibration accurate to 
0.006\,\AA\, across its effective energy range
(0.8--8.5\,keV/\,16--1.5\,\AA), approximately twice as good as that of
the MEG which has FWHM $\Delta \lambda \simeq 0.023$\,\AA\ 
and calibration accurate to 
0.011\AA\, over 0.5--7\,keV\,(25--2\,\AA).  
HEG has a higher effective area than
the MEG above about 6\,keV and is the only grating for which useful
data can be accumulated in the Fe\,K region. Specifically, HEG provides 
energy resolution FWHM\,$\simeq 40$\,eV, equivalent to a velocity resolution 
about 1900\,km\,s$^{-1}$  at 
6.4\,keV, a factor of 4 improvement over 
the {\it ASCA} CCDs' best period of performance.  The {\it Chandra}
 Low Energy Transmission Grating (LETG) is a single grating assembly
that complements the HETG by providing grating data to very soft
energies with $\Delta \lambda=0.05$\,\AA\ covering the 
0.07--6.0\,keV/175-2\,\AA\ regime.  Events dispersed by the gratings are
collected by a focal plane detector and can be assigned an energy
based on the position along the dispersion axis.  The  Advanced CCD Imaging Spectrometer (ACIS) 
comprises one focal plane 
instrument for {\it Chandra}.  ACIS has two CCD
configurations,  ACIS-S and ACIS-I optimized for spectroscopy and
imaging, respectively.  A micro-channel  High Resolution Counter 
(HRC) provides alternative focal plane instrumentation.  
While grating spectra can be dispersed onto either of the 
{\it Chandra} focal plane detectors, the usual configuration is to
disperse HETG onto  ACIS-S, whose CCDs are configured in a line
suitable for the dispersed spectrum.  Since the ACIS CCDs have 
intrinsic energy resolution, background events can be rejected with 
high efficiency, and different spectral orders can be discriminated
easily, making ACIS-S a highly desirable focal plane detector choice
for use in grating observations that do not require the very soft
response of the LETG. The LETG grating data are dispersed
onto the  HRC or ACIS detector. 
The  physical dimensions of the HRC can accommodate the
large dispersion suffered by the softest photons (photons below 
0.2\,keV are dispersed off the edge of the ACIS-S array).
 
\subsection{XMM-Newton}
The {\it XMM-Newton} observatory has a pair of gratings that comprise the 
Reflection Grating Spectrometer (RGS) \citep{denherder01a},
dispersing onto arrays of focal plane CCDs, providing a resolution FWHM\,$\simeq 2.9$\,eV at 1\,keV, or 
$\Delta\lambda \simeq 0.07$\,\AA\ . The absolute wavelength calibration for the RGS is 
0.008\AA\, equivalent to 100\,km\,s$^{-1}$ at about 0.5\,keV. {\it XMM-Newton} also carries both 
metal oxide semi-conductor (MOS) and pn-CCDs covering 0.5--10\,keV with 
energy resolution FWHM\,$\simeq 130$\,eV at 6\,keV, and FWHM\,$\simeq 55$\,eV at 1\,keV, 
with throughput comparable to the previous CCD instruments on board {\it ASCA}.  
Useful RGS data can be accumulated 
over the range 0.4--2.0\,keV/31--6\,{\AA}: data below 0.4\,keV are not used because of a
detector feature, and above 2.0\,keV the effective area is very low. 
Although offering a lower spectral
resolution than the {\it Chandra} gratings, RGS gratings offer
greater effective area over this range then either HETG 
or LETG.  

The choice of grating instrument for a given
scientific observation is a complex function of anticipated source
flux, spectrum and scientific objective. 
These grating detectors provide unprecedented resolution for many 
key diagnostic lines. Soft-band spectroscopy 
gained almost two orders of magnitude improvement in spectral resolution in the 0.5--1\,keV regime, 
with He-like triplets of O and Ne 
becoming resolvable.  Despite improvements in throughput,  it is still the finite signal-to-noise 
that provides the limiting factor in error determination for many astrophysical features. The 
potential of existing gratings is arguably best illustrated by the 900\,ks {\it HETG} exposure on 
NGC 3783, the deepest single 
AGN exposure to date with a grating detector (Figure~\ref{fig:3783lambda}); 
this dataset yielded more than 100 detections of absorption lines \citep{kaspi02a}.

\subsection{Suzaku}
Complementing the grating facilities is the Japanese-led {\it Suzaku} mission launched in 2005. 
\suzaku \citep{mitsuda07} carries four X-ray telescopes that each  focus X-rays on to a CCD 
forming part of the X-ray Imaging Spectrometer array (XIS) \citep{koyama07} . 
XIS CCDs 0, 2 and 3 are front-illuminated (FI) and  cover about $0.6-10.0$\,keV 
with energy resolution FWHM\,$\simeq 120-150$\,eV at 6\,keV. 
Use of XIS\,2 was discontinued after 2006 November because of a  charge leak that occurred.
XIS\,1 is a back-illuminated CCD with an enhanced soft-band response (down to 0.2\,keV) compared to 
the FI units, but  lower area at 6\,keV than the FI CCDs as well as a larger 
background level at high energies. 
\suzaku also carries a non-imaging, collimator  Hard X-ray Detector 
\citep[HXD,][]{takahashi07}
made from silicon PIN diodes, providing 
useful AGN data typically over 15--70\,keV, and a GSO well-type Phoswich counter, providing 
data over about 75--165\,keV for the handful of AGN  bright enough to be detected.  
The hard X-ray detector has a significantly lower background level than the previous {\it BeppoSAX} PDS 
instrument over most of its bandpass, and so the  simultaneous  XIS and PIN  data obtained 
during a {\it Suzaku} observation have allowed some exciting new insight into AGN.  

\suzaku also carries the now non-functioning  X-ray calorimeter XRS-2, this was the first 
X-ray calorimeter to  achieve orbit on a satellite. Calorimeters are 
highly desirable X-ray detectors because 
they can achieve excellent spectral resolution across a broad bandpass by  
converting the absorbed energy of incident X-ray photons into heat. 
The detectors are low heat-capacity absorbers, 
cooled to very low temperatures, about $0.1$\,K, and coupled to heat detectors (thermistors or
Transition Edge Sensors). 
XRS-2  would have provided both high resolution and high 
throughput in a single instrument with resolution better than  6\,eV across its bandpass.
Unfortunately a problem with the instrument vent configuration 
rendered XRS-2 inoperable after a few days in orbit. This failure followed a previous loss 
of the XRS-1 calorimeter along with the entire {\it ASTRO-E} satellite, which failed to achieve orbit 
after a problem with the first stage rocket. 

New results from {\it Chandra} and {\it XMM-Newton} and {\it Suzaku}  have 
opened up our understanding of AGN and we now review some of the recent developments and 
their implications. 

\section{Recent Observational Results}
\label{sec:recent}

\subsection{Absorption in a new regime}\label{newabsorption}

At the turn of the millennium, ionized ``warm absorbers'' had been established as
commonly occurring in AGN, with multiple zones detected across the
source population, covering a range of ionization and showing
variability on relatively short timescales (Section~\ref{sec:early}).
AGN spectra were revealed to be complex when observed at grating
resolution. As UV spectroscopy had already shown AGN to possess
kinematically-complex absorption, the richness of the X-ray absorbers
was no surprise.  In the longest grating exposures several ionization
zones of gas could be discerned in a single source \citep[e.g.][]{kaspi02a,steenbrugge05a}.  
In one case, NGC\,3783, \citet{netzer03a} and \citet{krongold03a} discussed the observed
complexity in the context of several phases of absorbing gas having
distinct temperatures and ionization states \citep[see also][]{behar03a}; this is the most common
approach to modeling AGN spectra.  In an alternative treatment,
\citet{goncalves06a} modeled the data using a single medium in total
pressure equilibrium; this showed a stratification of ionization
structure yielding multiple temperature components within the cloud
without a need to invoke separate absorber regions (although if zones
are found to be kinematically distinct such a picture would need to be
modified).  

Grating spectra allowed separation of lines that had been unresolved
using  CCDs.  Broader features such as radiative recombination
continua could be measured (or fit) for the gas temperature; the
components of some He-like triplets (O, Ne) could be separated and
component ratios used as a temperature and density diagnostic. The gas
temperatures indicated for the line-emitting gas were generally in the
million-K regime based on soft-band absorption lines.  An improved
ability to identify individual lines also led to better estimates of
gas kinematics, showing outflow velocities covering the range 
hundreds to thousands of km\,s$^{-1}$ \citep{kaspi02a,krongold03a,
blustin05a,mckernan07a,blustin07a}.  The detection of some unresolved transition arrays
\citep{sako01a,krongold05a,smith07a} and possible detection of dust absorption edges
\citep{lee01a} also implied the identification of cool and dusty zones of the absorber
complex, previously difficult to distinguish from absorption edges and
broadened emission lines.

\begin{figure}
\begin{center} 
\hbox{
\includegraphics[scale=0.35,angle=-90]{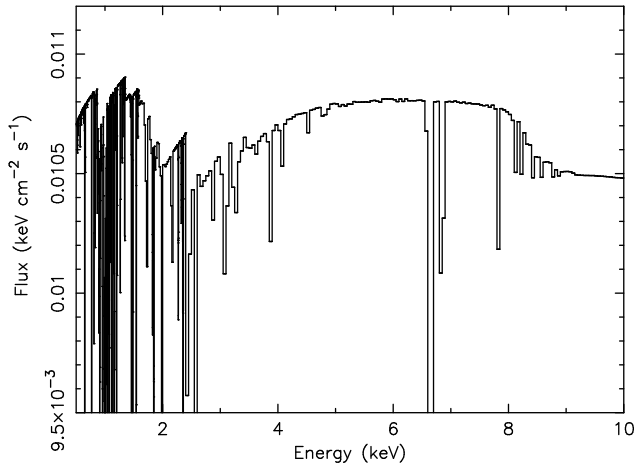}
}
%\vspace{-1.0cm}
\end{center}
\caption{\footnotesize
{Model of the high-ionization warm absorber fitted to the spectrum of
NGC\,3783 by \citet{reeves04a}.
The model line shows the imprint of a column of gas having 
$\log\xi \sim 3$ and ${\mathrm{N_H}} \sim 5 \times 10^{22}$\,cm$^{-2}$ 
over a smooth $\Gamma=2$ powerlaw continuum. 
The curvature  is due to high ionization K-shell edges and lines from 
Mg, Si, S, as well as from the L-shell and K-shell of Fe. 
{\em Figure reproduced from \citet{reeves04a} by permission of the AAS.}
}} 
\label{fig:3783opacity}
\end{figure}

In addition to the detailed diagnostics of known gas layers, 
new observations by {\it Chandra} and {\it XMM-Newton} revealed that
X-ray absorbers extend into a previously poorly-explored regime of 
high column density, high ionization gas. 
Deep K-shell absorption lines  were found in grating and CCD
spectra from very highly ionized species of Fe.  The detection of  
a strong absorption line from highly ionized Fe in a long
{\it XMM-Newton} observation of NGC~3783 \citep{reeves04a} confirmed the  suggestion of 
Fe {\sc xxv} absorption made in {\it Chandra} data \citep{kaspi02a}. 
The absorption line  showed 
variability in equivalent width over days, being strongest when the continuum level was
highest. In the case of NGC~3783 the absorption line was found to be
consistent with an origin in gas having $\log\xi \simeq 3$ and ${\mathrm{N_H}} \simeq 5 \times
10^{22}$\,cm$^{-2}$, likely existing about $0.1\,$pc from the nucleus.
An interesting consequence of the high-ionization warm gas is that a
large part of the curvature in the Fe\,K regime was accounted for by 
this zone, reducing the requirement for a relativistically blurred disk-line in the
source (Figure~\ref{fig:3783opacity}). 

Further examples of the presence of a  highly-ionized X-ray absorber showed up in other
long AGN exposures.  \citet{young05a} using {\it Chandra} HEG data,
and \citet{miniutti07a} in {\it Suzaku} data found lines at 6.7 and
6.9\,keV in MCG--6-30-15 most likely identified with Fe\,{\sc xxv} and
Fe\,{\sc xxvi}, with outflow velocity about $1800$\,km\,s$^{-1}$, supported
by detection of Si\,{\sc xiv} and S\,{\sc xvi} \citep{young05a}. 

In analysis of {\it XMM-Newton}  data from 
NGC~1365, \citet{risaliti05a} described four absorption lines in the
observed range 6.7--8.3\,keV, identified as Fe\,{\sc xxv} and Fe\,
{\sc xxvi} K$\alpha$ and K$\beta$ lines from gas outflowing at
1000--5000\,km\,s$^{-1}$.  This velocity range was interesting, being somewhat 
 higher than that typical for the lower-ionization gas 
detected in the soft-band.  \citet{risaliti05a} found the 
absorption lines to have high 
equivalent widths (about $100$\,eV) and these, combined with the
ratio of  strengths of the lines, implied the absorbing gas to have a
column density about $5 \times 10^{23} {\mathrm{cm}^{-2}}$ and 
to reside about 50--100\,\rg\ from the central source. In a follow-up 
campaign using {\it Chandra} the data showed variations consistent with an occultation 
by a Compton-thick cloud crossing the  line-of-sight of the X-ray source.  
In the context of the occultation model 
the source was estimated to be less than $10^{14}$\,cm in extent and 
residing within $10^{16}$\, cm of the nucleus  \citep{risaliti07a}. 

Mrk~766 also shows strong absorption features, observed at 6.9
and 7.2\,keV: in this case an explanation as Fe\,K$\alpha$ and
Fe\,{\sc xxvi}\,Ly$\alpha$ absorption is appealing, as a single outflow 
velocity of 13,000\,km\,s$^{-1}$ then fits both features
\citep{miller07a}.  Similarly to the case of NGC\,1365, a 
very high column density ${\mathrm{N_H}} > 10^{23} {\mathrm{cm}^{-2}}$ 
and high ionization, $\log\xi \ga 2$, was found for the absorbing gas.

\begin{figure}
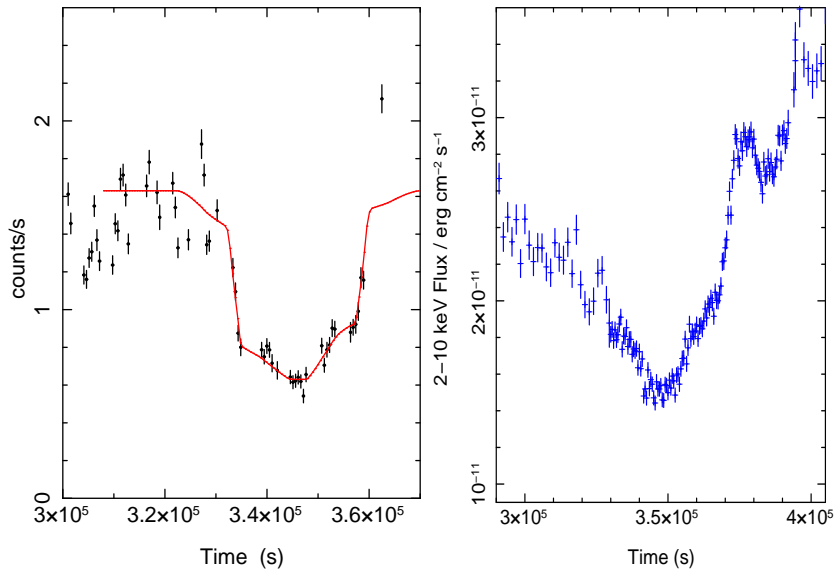

\begin{center} 
\hbox{
\includegraphics[width=55mm,height=75mm]{fig5a.eps}
\includegraphics[width=55mm,height=75mm]{fig5b.eps}
}
\end{center}
\caption{\footnotesize
{A close-up view of the deep dip section of the 
light curve of: left) MCG--6-30-15 
revealing the 
symmetric structure in the dip during  an 
{\it ASCA} observation from 1994 July. SIS0 data are shown, 
binned at 512\,s. The solid line is the predicted light curve based on 
a uniform cloud occulting a bright continuum point source situated 
within a ring of 
lower intensity extended emission. {\em Figure courtesy of Tahir Yaqoob, 
based on the 
analysis of  \citet{mckernan98a}}; right) NGC\,3516, 
showing the data discussed by \citet{turner08a}}  
} 
\label{fig:dips}
\end{figure}

Analyzing NGC\,3516, \citet{turner05a} discovered a  high-column absorber with  
${\mathrm{N_H}} \sim 10^{23} {\rm cm^{-2}}$ and log $\xi\sim 2$, 
covering about $50\%$ of the continuum source. 
Data from  a joint {\it XMM-Newton}/{\it Chandra} campaign during 2006 showed 
that changes in covering fraction of the high-column absorber could explain  the spectral
variability observed in this source \citep{turner08a}. Intriguingly, a 
deep dip in flux showed a complex but well-defined and symmetric flux profile 
that was similar to that 
observed in {\it ASCA} data from MCG--6-30-15 \citep[][Figure~\ref{fig:dips}]{mckernan98a}.  
The dip profiles could be explained as an eclipse of the
continuum by a cloud moving across the sight-line \citep{mckernan98a,turner08a}. 
If an occultation origin for such deep dips
could be proved then the potential would be enormous: the angular size of
black hole event horizons is $\sim 10^{-6}$\,arcsec for the nearest AGN, 
so imaging of these systems is not possible in the foreseeable future,
however, using the occultation event in MCG--6-30-15,
\citet{mckernan98a} effectively mapped the nuclear regions on a size
scale of $\sim 10^{-7}$\,arcsec.  Such angular scales may otherwise
only be probed by analysis of micro-lensing 
events \citep{chartas04a, chartas08a}.
For NGC 3516, the {\it XMM-Newton} data showed that an 
additional gas zone was evident at a similar column density to that responsible for the 
spectral variability, but with higher ionization, log $\xi \sim 4$, producing deep 
absorption lines from the K shell of Fe\,{\sc xxv} and Fe\,{\sc xxvi} ions \citep{turner08a}. 

As existing AGN observations cover a wide range of signal-to-noise ratio 
it is too early to make a definitive assessment of the occurrence rate of these  
high column X-ray absorbers with  ${\mathrm{N_H}} \sim 10^{23} - 10^{24} {\mathrm{cm}^{-2}}$ 
and having  Fe\,{\sc xxv} and Fe\,{\sc xxvi} as the dominant ions.  However, as these 
lines have shown up in a number of well-exposed  AGN observations,
the gas, with typical outflow velocities of a few to many thousand km\,s$^{-1}$, 
appears common in local AGN. 

In addition to the strong observational effects attributed to changes
in gas covering and/or opacity, any circumnuclear gas must respond to
changes in its local ionizing continuum. Whether ionization changes
can be observed depends on not only the signal-to-noise of the data
but also whether the gas lies in a regime where changes in the
illuminating continuum produce an observable change in the gas
state. Some results have been published in this regard, with
\citet{kraemer05a} finding changes in the X-ray absorbers of NGC\,4151 
consistent with a response of the gas to changes in the ionizing
continuum. Also, \citet{netzer02a} found evidence for a response of
the absorbing gas to continuum variations in NGC\,3516 \citep[see also][]{netzer03a}. 

Some sources show evidence for extremely  high velocities in the outflowing gas. 
Outflows seem often to be associated with the more luminous AGN, and
typically have velocities in the range 0.1--0.2~c \citep{reeves08a}.
{\it RXTE} data for PDS~456  first showed a very deep Fe\,K  edge revealing the existence of a 
column of gas with  
$\mathrm{N_H} \sim 5 \times 10^{23} {\rm cm^{-2}}$ and $\log\xi \sim 2.5$ in the source. 
A later  {\it XMM-Newton}  observation allowed a determination of an outflow velocity 
of 50,000\,km\,s$^{-1}$ \citep{reeves03a}. Most recently, 
a long \suzaku observation has resolved the deep absorption feature 
into absorption lines from Fe\,{\sc xxv} and {\sc xxvi} 
yielding a revised estimate for the flow velocity of 0.25--0.3\,c.
Other early reports of such features included 
APM~08279$+5255$ \citep{chartas02a}, PG~1115$+80$ \citep{chartas03a}, 
PG~1211$+143$ \citep{pounds03a}, 
IRAS~13197--1627 \citep{dadina04a}, Mrk~509 \citep{dadina05a}, IC~4329A
\citep{markowitz06a} and MCG--5-23-16 \citep{braito07a}. 
Further to these outflows,  high-velocity inflows have been suggested to explain 
apparent redshifted  resonance
absorption lines detected in some sources 
(e.g. Mrk~335, \citealt{longinotti07a}; Mrk~509, \citealt{dadina05a};
E1821$+643$, \citealt{yaqoob05a}) 
with velocities in the range 0.1--0.4~c \citep[see][and references
within]{reeves08a}.

Not just restricted to the X-ray data, high velocity outflows are a well-known 
general phenomenon in AGN, being observed unambiguously in optical and ultra-violet spectra, 
especially in broad absorption line quasars
\citep[e.g.][]{weymann91a}.  The general existence of significant
outflows is also implied by the existence of jets in AGN
\citep[e.g.][]{axon89a}, although we likely need to make a
distinction between the highly collimated jet outflows and possible
wind-type outflows with wide opening angles \citep[e.g.][]{pounds08a}.
However, the X-ray results remain contentious, with some authors  claiming 
a possible correlation between 
the cosmological recession velocity of the AGN and the absorption
outflow velocity for a number of sources \citep{mckernan05a},
suggesting the absorbers have an origin in local gas, particularly
since some of the sources in question lie behind the local hot bubble
of gas known as the Northern Polar Spur. Countering these arguments,
\citet{reeves08a} calculate that the variability observed in PG1211$+143$ over 4
years shows the absorber to be too compact, and the surface
brightness of the gas too high, for the observed absorption signature
to arise in local gas; furthermore, \citet{reeves08a} contend that the
velocity coincidence observed in PG~1211$+143$ does not extend to
other AGN. Finally, claims have been made that some of the detections
of narrow absorption (and emission) lines are not statistically
significant and we return to that issue in
Section\,\ref{narrow_emission_lines}.

\subsection{Broad-band spectral variability} 
While the existence of spectral curvature in the $2-8$\,keV regime, 
dubbed the ``red wing'', was confirmed by
{\it XMM-Newton}, {\it Suzaku} and {\it Chandra} data
(\citealt{wilms01a}, \citealt{vaughan04a}, \citealt{miniutti07a},
\citealt{nandra07a}), 
it has continued to  prove difficult to make a convincing distinction
between models dominated by either complex absorption or blurred reflection to
explain how that curvature arises  (c.f. \citealt{miniutti07a},
\citealt{miller08a}). 

\begin{figure}
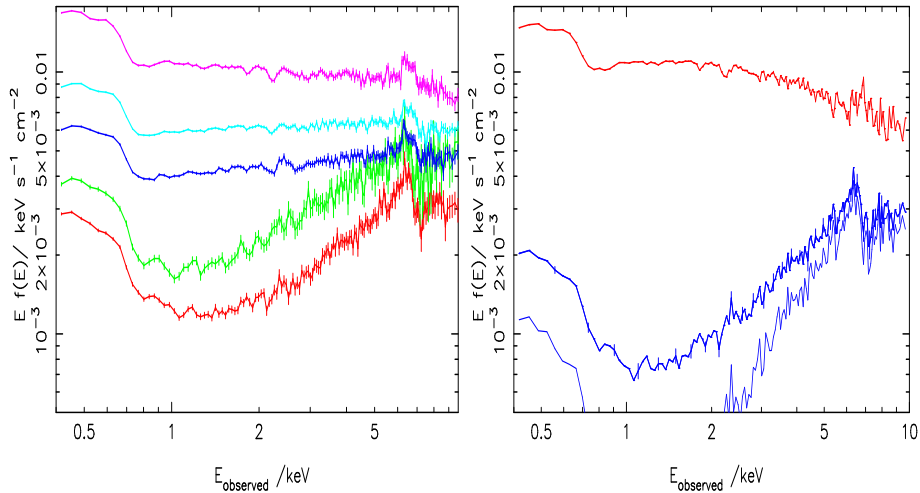

\begin{center} 
\hbox{
\includegraphics[width=65mm,height=60mm,angle=-90]{fig6a.eps}
\includegraphics[width=65mm,height=60mm,angle=-90]{fig6b.eps}
}
\end{center}
\caption{\footnotesize
{Left: Data used for the principal components analysis 
of Mrk 766, from {\em XMM-Newton}, averaged into five 
logarithmically spaced flux 
states based upon the 1--2\,keV\,flux, and ratio-ed to a power-law of photon index
$\Gamma=2$ with unit normalization,
illustrating how the source hardens 
to low fluxes, behavior typical of Seyfert galaxies.  Based upon data 
and analysis presented by \citet{miller07a}. 
Right: Principal component spectra of Mrk 766, 0.4--10\,keV. 
The first, varying, principal component is the upper line (red), 
the possible range of an unvarying zero-point spectrum is shown by the lower set of 
spectra (blue). For clarity, error bars are only plotted on every fifth spectral point
on both panels.
}} 
\label{fig:766}
\end{figure}

The most progress has come from consideration of the 
spectral changes that occur in these sources  over timescales as short as tens of 
ks \citep[e.g.][Figure~\ref{fig:766}]{miller07a}. Local AGN commonly show a 
systematic hardening as a source dims to a low flux level  
\citep[e.g.][]{papadakis02,pounds04a, pounds04b,vaughan04a, miller07a,miller08a,turner08a}. 
As significant spectral variability occurs on much shorter timescales than 
a typical observation duration, the  commonly-fitted mean spectrum 
is actually a superposition of source spectra from
differing states.  Fitting to mean spectra alone misses key 
information.  Worse, if source variations are non-additive, such as in the
case of variations caused by varying opacity, then fitting to the mean spectrum
is unlikely to give a good representation of the physical construction of these systems.  
 
A simple method to address the problem is to 
take time- or intensity-selected spectra to try to isolate the
physical origin of the spectral variations.  When sufficient photons
have been accumulated, one can examine the flux correlations at
different energies and look for the spectral forms of the smallest
number of spectral components that can be used to describe the data.

More versatile than time-resolved spectroscopy, a powerful
mathematical tool used to analyze systematic variations is principal
components analysis (PCA). Its use in analysis of AGN X-ray spectra,
albeit with low energy resolution, was first demonstrated by
\citet{vaughan04a}, who showed that MCG--6-30-15 could be described by
a constant hard component and a variable softer one. The steady
spectral component carries the so-called red wing, showing that component to be
surprisingly constant in amplitude while the observed continuum flux
varies, contrary to the line/continuum correlation that would be
expected if the line is produced very close to the continuum source.
This behavior has been addressed with models that invoke variable
light-bending effects driven by variations in the height of the
continuum source above the disk \citep{fabian03a, miniutti03a,
miniutti04a}. In the light-bending model, as the continuum source
moves closer to the disk and black hole, general relativistic effects become more
pronounced and some of the photons that were previously able to reach
the observer instead arrive at either the disk surface or the black
hole.  For appropriate choices of height and range in height of the source,
the fractional change in the primary source intensity viewed by
the observer can be significant with the fractional change in the disk
reflection intensity remaining relatively small \citep{miniutti03a}.  In this
model, the continuum variations seen by the observer are thus
primarily due to the effect of the moving source of X-ray continuum.
For the effect to be isotropic, a ring-like structure for the illuminating
source could be envisaged \citep{miniutti03a}.
The model has been applied to explain the variable spectra of a number
of sources, including MCG--6-30-15 \citep{miniutti03a,
miniutti07a}. The interpretation of the observed hard spectral
component as blurred reflection requires significant contributions
from within 6\,\rg\ in order to obtain a component that extends to low enough
energies to fit the data, a result that has been used to infer that the
black hole in MCG--6-30-15 is spinning \citep[e.g.][]{brenneman06a}.

\citet{miller07a} address the problem in standard PCA applied to X-ray spectra where the
number of data slices (often limited by the requirement of having adequate signal-to-noise 
in each data slice) is less
than the number of energy intervals (determined by instrument energy
resolution) and consequently the covariance matrix is singular and the
eigenvectors are not uniquely defined.  In this case the leading
eigenvectors may be extracted using Singular Value Decomposition (SVD)
\citep{press92a} and employment of this technique led to extraction of
the leading eigenvectors at the full instrument resolution.  
Application of PCA
with SVD was performed on Mrk\,766 \citep{miller07a} yielding a
spectral decomposition into a variable amplitude power-law of slope
$\Gamma \simeq 2.4$ with a modestly broadened ionized component of Fe\,K
emission superposed on this (Figure~\ref{fig:766}).  The data showed a correlation of the flux
in the continuum and Fe line with no detectable lag \citep{miller06a}.
This result was supported by apparent variations in the peak
energy of the ionized line component. The energy variations, lack of
lag and the width of the ionized emission line are all consistent with
the line being produced at around 100\,\rg.  
Another line response to continuum was seen 
as a one-time event in  MCG--6-30-15 where \citet{ponti04a} found evidence for a 
Fe\,K$\alpha$ component that appeared 3000\,s after one strong continuum flare,
suggesting some contributions to the observed emission line may come from
regions located close to the continuum source ($< 240$\,\rg) but
outside the regime where light-bending effects are significant.
In the analysis of Mrk\,766, \citet{miller07a}
found in addition a hard-spectrum, less time-variable component that
could be fit either as a blurred reflector, unblurred reflection
from an ionized wind or absorption from a disk wind whose variable
covering of the power-law explained the spectral variability observed. While 
the decomposition of X-ray data had isolated the variable spectral component, 
the interpretation of this `steady' component was still unclear. 

Turning to analysis of  
the well-studied source MCG--6-30-15, \citet{miller08a} compiled
all the best long-exposure, high-quality data, obtained from
522\,ks with {\it Chandra} HETG, 282\,ks with {\it XMM-Newton}
EPIC-pn/RGS and 253\,ks with {\it Suzaku} XIS/PIN and decomposed the spectral
variations using the SVD-PCA.
The \citeauthor{miller08a} approach was to make the spectral decomposition and then  fit
the variable and steady components as for Mrk~766.  The model for the
variable component was found to be a simple power-law $\Gamma \simeq 2.3$
covered by the complex warm absorber.  The warm absorber in
MCG--6-30-15 has long been known to consist of at least three zones of
ionized gas whose signatures are seen in grating observations
\citep{lee01a, turner03a, turner04b}.  
In modeling the hard component it was found
that a variable partial-covering zone of absorption plus absorbed
low-ionization reflection (distant from the source) resulted in 
a full model that fits all flux states of the
data over its entire observed energy range.  
The construction of this model was different to that of
\citet{young05a} in allowing some of the ionized absorption to have a covering
fraction $<1$: the spectral shape is then reproduced by gas of intermediate
ionization, $\log\xi \la 2$, and
consequently the  absorption model of \citeauthor{miller08a} does not predict the
resonance Fe\,K$\alpha$ absorption lines at 6.5\,keV that \citeauthor{young05a} had previously
suggested would be expected if absorption does cause the red wing.
In the \citeauthor{miller08a} model, the relative lack of variation of the red wing is an artifact
of variation in the partial covering fraction of the absorber.

Seyfert behavior  is sometimes complicated by the observation of spectral changes
{\it within} the lowest flux states.  
\citet{reynolds04a} interpreted spectral variations during the low state
for MCG--6-30-15 as correlated changes between the line and continuum
and explained the lack of correlated behavior at higher fluxes as due
to a saturation of line production beyond a critical continuum
level. It has been suggested that such complex low-state  behavior could be
expected either from patchy ionization of the disk surface
\citep{reynolds00a} or in light-bending models if the continuum source
is located close to the axis of spin for the black hole
\citep{miniutti04a}.  In
the alternative absorption models, behavior within the low-states 
may be explained either as opacity or ionization changes in the absorbers
or as systematic
evolution in the relative sizes of the source and the absorbing structure
\citep{miller08a}.

\subsection{Hard X-ray spectra}\label{sec:hardspectra} 

Interestingly, fluxes observed above 10\,keV often exceed 
those expected from reflection  from the surface of a disk subtending
$2\pi$ steradians to a continuum source. Relative reflection strengths  in 
the range $2 \la R \la 4.8$ are inferred 
in MCG--6-30-15, (e.g. \citealt{ballantyne03a,miniutti07a}).
\citet{ballantyne03a} proposed a model of double reflection to explain
both the spectrum around the Fe\,K$\alpha$ line and the high $R$ value.
The light-bending model, by design, enhances the reflected
component over that expected in the absence of relativistic
aberration and thus is constructed to produce enhanced $R$ values. 
Other explanations for the
relative lack of variability of the ``red wing''
(e.g. \citealt{merloni06a}, \citealt{nayakshin02a}, \citealt{zycki04a})
do not naturally explain the high $R$ values, although it is possible
that deeply-embedded emission hotspots in an inhomogeneous disk might
be able to produce such an effect. 

For NGC\,4051, reflection models yielded 
$R \sim 7$ \citep{terashima08a}. Light-bending models appear not to work for this source as 
they cannot simultaneously account for the high R value and the weak 
Fe\,K$\alpha$ emission, leading \citeauthor{terashima08a} to conclude that 
partial-covering absorption is required.  
Attempts to model the spectrum of Mrk~335 also required 
$R\sim 2.8$ \citep{larsson08a}. 

An observed feature of the hard-band flux is that it appears less variable than the
2--10\,keV flux \citep{miniutti07a}, consistent with having a joint origin with
the steady red wing.  In the blurred reflection explanation this is consistent
with the  light-bending model.  In absorption-dominated
models the inference would be that much of the 2--10\,keV band variability is
caused by variations in absorption, with the steady continuum source being 
relatively unaffected by absorption above 10\,keV. 

A \suzaku observation of 1H\,0419--577 appears to  provide a rare case of a 
distinction between models.  A marked `hard excess' of counts was
detected in the PIN data relative to the predicted flux based on model
fits below 10\,keV. The \suzaku data can be fit using an
absorption-dominated model but not with any reasonable blurred
reflection model. The detailed fits  show that Compton-thick partial-covering gas
must exist in 1H\,0419--577 (Turner et al. 2009, ApJ in press). This result 
supports the earlier contention that partial-covering shapes the observed spectrum
and much of the flux variability in 1H\,0419--577 
\citep{page02,pounds04a,pounds04b}.  Considering the broad-band properties of 
1H\,0419--577, the source is 
consistent with spectral modification by a clumpy disk wind that provides the X-ray absorption
from gas residing at radii inside or comparable to the radius of the
optical/UV BLR. Importantly, the observed 
Fe\,K$\alpha$ line luminosity is consistent with an origin in an
equatorial disk wind in that case (Turner et al. 2009).  

The luminous AGN PDS~456 also exhibits a hard excess (at the 3$\sigma$ level) 
that similarly can be explained only by
partial-covering models (Reeves et al. 2009, submitted). These two
sources provide the first strong evidence for Compton-thick
partial-covering gas in type\,1 AGN and show that the intrinsic luminosity of 
such sources is underestimated  significantly when
based on data below 10\,keV.  The {\it Swift} BAT survey has discovered 
many new AGN  based on their flux above 10\,keV, suggesting that there 
may be a much higher fraction of heavily absorbed sources in the AGN 
population than previously known \citep{tueller08}. Further study of these 
absorbed AGN should give us a much better understanding of the  
properties of the absorbing gas in accreting systems. 

\subsection{Origin of the soft excess}

The soft X-ray spectra of AGN often show a marked rise in flux below about
1\,keV, the so-called `soft excess', relative to the downwards
extrapolation of the higher energy spectrum
\citep{arnaud85a,turner89a}.  Such excesses are particularly prevalent in
narrow-line Seyfert\,1 galaxies \citep{boller96a}.
While AGN have a soft-band contribution
from extended emission (binaries, hot gas and in Seyfert type\,2 a
few percent of scattered nuclear radiation), this is observed at a low
flux level and typically comprises only a few percent of the soft-band
flux: the soft excess discussed here is not dominated by those extended 
components.

Historically, the soft excess has often been fit using a black body
model, yielding best-fit temperatures in the range 0.1--0.2\,keV
\citep{walter93a,czerny03a}.  Although the temperature of the soft excess is too high
to be direct emission from the inner disk (unless photon trapping is
invoked, see \citealt{abramowicz88a,mineshige00a}), it could represent
Compton-scattered disk photons
\citep[e.g.][]{czerny87a}. However, \citet{gierlinski04a} fit models
representing a Compton-scattered disk component to a sample of AGN,
finding a surprisingly narrow range of temperatures for sources
covering such a large range ($10^6-10^8 {\mathrm M}_{\odot}$) in mass.

Grating data have clearly shown strong emission and absorption
features in the soft X-ray spectra of most AGN, these arise in the
warm absorber and other circumnuclear reprocessors in and out of the
line-of-sight.  As discussed by \citet{gierlinski04a}, the imprint of
strong spectral features such as O\,{\sc vii}, {\sc viii} and the Fe
M-shell unresolved transition array (UTA) in the 0.7\,keV regime could
affect the perceived shape of the soft excess.  The problem with
assigning an `atomic interpretation' to explain the similarities of
soft excesses across the Seyfert population is that some sections of
the soft excess often appear smooth and certain predicted features are
not evident.  This led \citet{gierlinski04a} to consider the
possibility of relativistic blurring of absorption features arising in
a relativistic outflow, whose reduced opacity in the soft X-ray regime
(owing to the high ionization of the gas) would then explain the spectral soft
excesses. In the context of such a picture, \citet{schurch07a,
schurch08a} computed the X-ray spectra of columns of outflowing gas
as functions of density and velocity, demonstrating that very high
outflow velocities would be required if absorption in outflows are
solely to explain soft excesses observed in AGN spectra.  The high
velocities at about 0.9\,c exceed even the relativistic wind components
detected through energy-shifted absorption lines, and
\citet{schurch08b} have found that models of line-driven accretion-disk winds
do not attain sufficiently high velocities.  \citet{schurch08a}
note that magnetic driving is likely the only acceleration mechanism
that could achieve the high velocities required in this case, and that
the wind itself would have to be clumpy, only partially covering the
source.

\citet{chevallier06a} assert that a blurred absorber would have to be
in pressure equilibrium, otherwise the predicted spectral variability
exhibited by a source and across a sample would be much larger than
that observed.  \citet{done07a}, however, have disagreed with this
conclusion.  A relativistically blurred absorber would have associated
relativistically-blurred line emission, that might dominate in sources
viewed from certain angles. Models composed of blurred emission lines
from an ionized reflector have been invoked to explain some RGS
spectra \citep{branduardi-raymont01a}, although that result was
challenged by \citet{lee01a} and \citet{turner03a, turner04b} 
who find that the soft X-ray spectrum
of MCG--6-30-15 may, instead, be influenced by the presence of a dusty warm absorber.

Interestingly, detailed analysis of Mrk~766 and MCG--6-30-15 found the
soft excess in those cases to be satisfactorily accounted for by the
combined opacity profile of several layers of unblurred absorption
\citep{turner07a,miller08a}. It may be that a complex of ionized
absorption may yet turn out to be the dominant cause of apparent
`soft excesses and that extreme velocities and other effects are
not required to explain these sources \citep{nicastro99a},
with only the extreme soft-excess AGN 
such as RE\,J1034$+396$ requiring an accretion disk 
contribution \citep[e.g.][]{pounds95a}.

Other attempts \citep{crummy06a} to explain the soft excess of AGN
include modeling the broad X-ray spectra using blurred, ionized
reflection models.  As discussed by
\citet{sobolewska07a}, a reflection origin strongly constrains the
strength of the soft excess relative to the incident continuum and the
observation of several extremely strong soft excesses would require
anisotropy of the continuum or light-bending, such that the disk {\em
consistently} sees more of the continuum radiation than the observer
in those sources.
In contrast, the absorption-based models fit the data without recourse to 
any radiation anisotropy.
Taking an overview of models, \citet{middleton07a} find that
absorption-based models for the soft excess provide a clearer
correspondence with black holes in the stellar mass class, compared to
reflection models.  Aside from the compelling nature of such an
extrapolation, how might the issue be resolved?  If the general lack
of variability of the Fe-line `red wing' is caused by light bending, one
should expect a blurred reflection component of the soft excess to
show similar non-variable behavior.  More complete studies of the
variability of the soft excess would then provide this critical test
of the reflection hypothesis.

\subsection{Variable narrow emission lines}\label{narrow_emission_lines}

While the evidence for a relativistically-broadened line is
contentious, the existence of a narrow `core' component of Fe emission
is common among observed local AGN. The energy of this core component is generally
consistent with 6.4\,keV \citep{yaqoob04a} implying an origin in
low-ionization material. Line equivalent widths lie in the range a few
tens of eV up to about 200\,eV
\citep[e.g.][]{nandra97a,sulentic98a,reeves02a}, allowing some
constraints to be placed on the covering fraction of the emitting
cloud (which must be high) and the optical depth of the emitting gas
\citep{leahy93a}.  The observed widths of the narrow line
`core' components are typically a few thousand km\,s$^{-1}$,
consistent with previous suggestions of an origin in the BLR
\citep{yaqoob01a,kaspi01a,yaqoob04a,bianchi08a} 
although some contribution from the material at large distances
is likely \citep[e.g.][]{reeves07a}.  
A study by \citet{nandra06a}
showed there to be no correlation between the width of the narrow Fe\,K$\alpha$
emission component and either H$\beta$ width or black hole mass; that result
suggested significant contributions to the narrow line may arise from
regions other than the optical BLR, including the 
outer parts of the accretion disk. A disk contribution to the narrow
component had also been suggested from study of
MCG--6-30-15 {\it Chandra} data \citep{lee02a}.

Additional evidence for narrow lines originating in the accretion disk
came from the observation of variations in the narrow Fe\,K$\alpha$ line in
Mrk~841 \citep{petrucci02a} and the discovery of narrow and rapidly
(tens of ks) variable lines in NGC~3516 \citep{turner02a}.  In the
case of NGC~3516 the lines appeared at energies redward of the neutral
Fe\,K$\alpha$ line and were fully separated in the data from the narrow line core
at 6.4\,keV; considerations of those line energies and strengths led
to a conclusion that they were likely Doppler-shifted Fe lines,
originating in the accretion disk or in an outflow, such as a disk
wind \citep{turner02a,turner04a}. Further examples of the phenomenon,
dubbed `transient Fe lines' soon came from observations of other AGN
\citep[e.g.][]{yaqoob03a,guainazzi03a,longinotti04a}.

The periodic pattern of line-energy/profile changes expected from a
hotspot orbiting a black hole are very distinctive, so long as 
the emitting gas is not  viewed 
too close to the polar axis (i.e. so long 
as there is a significant velocity component along the line-of-sight).  The line
evolution with azimuthal angle (around the disk) is particularly
distinctive for lines originating within 20\,\rg\ where general
relativistic effects are measurable
\citep{dovciak04a, czerny04a, goosmann07a, dovciak08a}.
Lines originating in a wind
would also suffer relativistic and Doppler effects related to motion
along the sight-line, but these are different to the periodic energy
shifts due to disk rotation.  Both possibilities are interesting:
confirmation of a disk origin would allow derivation of parameters
such as the radius of emission and inclination of the system,
confirmation of a wind origin might allow us to trace the velocity 
distribution of
emitting knots to help determine launch radius, acceleration/deceleration
mechanisms and mass loss rate.  In principle, one could confirm a line
origin using time-resolved spectroscopy with available
instrumentation; in practice, sufficient signal has to be accumulated
in distinct azimuthally selected slices to see the line energy
evolution, without integrating so long that the azimuthal variations are
averaged out.  Few observations to date have had the combination of
sufficient source brightness, hotspot radial location and observation
duration to track line evolution in any useful way and so the origin
of these features has been poorly tested.  \citet{iwasawa04a} claim
tentative evidence for periodic flux variability for a line in
NGC~3516, although only viewing 3 cycles of repetition.  Perhaps the
most positive development in this area was the previously noted
discovery of a line/continuum flux correlation in Mrk~766 \citep{miller06a} 
with associated changes in line energy  \citep{turner06a}.   This
flux correlation and peak-energy variability was discovered in data from a 500\,ks {\it
XMM-Newton} observation and would not have been significantly
detected in a shorter observation. 

Given the large number of claims of energy-shifted absorption and
emission features now in the literature, \citet{vaughan08a} have
questioned the significance of the mass of results reported. Reviewing 38 published 
detections of features having energy shifts $v \ge 0.05$c,
\citeauthor{vaughan08a} find a tight linear relationship between the
estimated feature strength and its uncertainty, demonstrating
that more accurate data tend to show weaker lines, whereas if
these are true features then the stronger lines should show up in a
well-defined way with improved signal-to-noise. The inference is 
that many of the reported detections are
merely statistical fluctuations. However, a number of these features
have been detected on multiple occasions in the same source,
e.g. features in NGC~3516 show up repeatedly in the 5--6\,keV band
(\citealt{turner02a,bianchi04a,dovciak04a,iwasawa04a}) and others have
been assessed using Monte Carlo simulations
\citep[e.g.][]{yaqoob05a}.  
Further long observations of AGN detecting  characteristic changes in 
line energy or flux with time, coupled with a rigorous statistical analysis, 
are required to settle the issue.

\section{Summary of status and a review of model-space}\label{sec:currentmodels}

That the X-ray spectra of AGN show marked curvature in the 2--8\,keV regime is 
now well-established and in itself, no longer controversial \citep[e.g.][]{nandra07a}, 
although the interpretation of this curvature is still not agreed upon.
Taking the ensemble of results both from sample studies and across the
literature, summarized above, it appears that models dominated either by
blurred reflection or by complex absorption provide
statistically-comparable fits to the {\em mean} X-ray spectra of  most AGN.  
We now summarize the points for and against the hypotheses that one or other
process dominates, before discussing more general models, arguably more realistic,
in section\,\ref{complexmodels}. 

In the following section we concentrate on ``reprocessing'' origins for spectral
signatures.  It should be born in mind however that we still have no established
model for the continuum production, and that process itself may imprint spectral
signatures and spectral variability on the observed spectra.  In particular
the corona-plus-disk emission models of \citet{haardt91a, haardt93a} suppose a coupling between
photons from the accretion disk and the hot Comptonizing corona, acting as the
primary cooling source, and between the Comptonized photons and the accretion disk,
acting as a heating source.  In this picture, 
the illuminating spectrum should not be considered as being a component
independent of the reflection or absorption arising from cooler material in the accretion disk.

\subsection{Blurred reflection}

The approach of many teams to modeling the curvature observed over the 
2--8\,keV band  is to use blurred reflection models 
obtained by convolving reflection spectra from \citet{ross05a} with
the effects of emission from a rotating disk deep in a black hole's
potential well \citep[c.f.][]{laor91a}.  First, any obvious absorbing gas is isolated by
measurement of narrow features in grating data \citep[e.g.][]{wilms01a}. 
The absorbing gas is
thus parameterized and model components or tables included in the fit
to allow for that; it is common to then  assume a
power-law continuum and blurred reflector to model the rest of the
spectrum. We assess the motivation for and implications of modeling
using this approach.

\subsubsection{Motivation}
One of the compelling aspects of blurred reflection models is that the
signatures from material within a few tens of gravitational radii 
might reasonably be expected to be produced in AGN
at a level that could be measured with existing
instrumentation.  If the X-ray continuum is produced on size-scales of
a few \rg\ then for certain disk geometries one would expect to see 
significant reflection component contributions from within 20\,\rg\ 
and these
should show measurable general relativistic effects. However,
the integrated spectrum from disk reflection depends on the
illumination pattern across the accretion disk and thus
depends on the size, geometry and location of the continuum source
and the structure and ionization of the accretion disk.

\subsubsection{Continuum size}
\label{continuumsize}
Production of the X-ray continuum over
an extended corona would not result in disk reflection that was
strongly weighted towards small radii and consequently one would not
expect to observe significant blurring in the integrated disk
reflection spectrum.
The size of the X-ray continuum source is therefore a key issue in
predicting the integrated reflection spectrum and assessing the
consistency of the data with that prediction.  One intriguing and
potentially strong constraint is the result of observations of
X-ray variability in gravitationally lensed images of quasar RXJ\,1131-1231,
which, if interpreted as due to micro-lensing variations
suggests the continuum region to have
half-light radius $\la 6$\rg\ \citep{chartas08a}.  

The case has long been made that the rapid variability observed in X-ray 
light curves  of AGN demonstrates, from arguments based upon  light-crossing timescales, that the 
X-ray emitter must be very compact \citep{rees77}: variations observed 
with current missions would suggest the size to be on the order of a thousand light seconds 
across. It is now unclear however, whether such rapid variability 
should be interpreted in terms of absorption variations.  For example, 
constraints on the continuum source
size have estimated the region to be a few \rg\ in extent based on
what appear to be (de-)occultation events in AGN \citep[e.g. NGC\,1365][]{risaliti07a}.
Whichever way one views the data, there are reasons to think the illumination of the disk
might indeed be centrally concentrated and thus that there could be significant
reflection within 20\,\rg\ if the accretion disk also extends that close to the black hole.

\subsubsection{Disk parameters}
The most recent systematic study of local AGN was that of  \citet{nandra07a}
based on available {\it XMM-Newton} data. It was concluded that a large fraction 
of local AGN show emission that appeared best-fitted with a blurred reflection model, but with a
large dispersion in the characteristic radius of emission.  However, the 
study did not allow the X-ray absorbers to have 
covering fractions less than unity when absorption models were
tested, nor was it able to test models against the spectral variability of AGN.
Once covering fraction is allowed as a free parameter then
absorption-based models are also acceptable fits to the
time-averaged X-ray spectra of AGN 
(e.g. Mrk\,766, c.f. \citealt{nandra07a}, \citealt{miller07a,turner07a}). 

\subsubsection{Variability}
Perhaps the biggest problem for the blurred-reflection-dominated
picture is the lack of any simple correlation observed between
continuum and reflection.   Light bending models rely on the continuum flux 
being intrinsically steady and  then explain the
observed phenomena by invoking changes in the height of the continuum
source above the disk, opponents of this picture argue that such
variation in continuum placement is {\em ad hoc} and might require
fine-tuning if it is observed that there is universally no
line-continuum correlation in samples of AGN.  Some other models seek
to explain the observed constancy of the red wing
\citep[e.g.][]{merloni06a,nayakshin02a,zycki04a} but these may have
difficulty reproducing the high apparent R values observed above
10\,keV.

\subsubsection{Strength of the reflection component}
Analysis of AGN over the broad bandpass afforded by {\it Suzaku} has
shown that some individual sources have very hard spectra indicating
nominal solid angles much greater than 2$\pi$  for blurred reflection
(Section~\ref{sec:hardspectra}).  
High geometrical factors may be
feasible if the disk has a deep funnel or other favorable inner
geometry, but as always,  the 
strength of the reflected spectrum depends critically on the 
surface illumination pattern, illumination angle and viewing angle. 
Another possibility for producing a high observed R value 
is that the illuminating continuum  is partly hidden from direct view and so what 
we see is only the  Thomson-scattered radiation (i.e. the reflector sees 
some continuum that is hidden from the observer, such that the observed  
ratio of continuum to reflection is not directly interpretable).  Some hard excesses
are sufficiently extreme that model fitting disfavors reflection with all reasonable 
modifications  (1H\,0419--577, Turner et al. 2009; 
PDS\,456, Reeves 
et al 2009, submitted).

\subsubsection{The question of spin}

In principle, if one has a clean view of the reflection spectrum then 
the extent of the `red wing' could constrain the inner radius of
the emission and thus the black hole spin: if it is found that
emission from radii $r < 6$\,\rg\ is required in the reflection
model then one may interpret the data as favoring an innermost stable
circular orbit $r_{\mathrm{ISCO}} < 6$\,\rg\ and thus indicative of a Kerr metric.
Claims for detection of black hole spin 
have been made in MCG--6-30-15 and IRAS\,18325--5926
\citep{iwasawa04b} where curvature is evident down to $\sim 2$\,keV. 
With the uncertainty and debate on the physical processes that dominate the X-ray
spectrum in this energy range \citep[e.g.][]{miller08a}, in the view of these authors it is not
currently possible to constrain the black hole spin in a model-independent
way.

\subsection{Absorption}
\label{sec:absn}

An alternative view  is that spectral signatures from complex
absorption dominate the X-ray spectra and may explain the spectral
variability of AGN.

\subsubsection{Motivation}

High-resolution UV spectroscopy shows that multi-layered, complex
absorbers are common in AGN  comprising low-column,
low-ionization systems, often showing several kinematic components 
\citep{crenshaw03a}.
The highest ionization UV emission lines in high accretion-rate
objects show systematic asymmetry \citep[e.g.][]{gaskell82a,
richards02a} supporting a disk wind origin for the UV absorbers
\citep{murray98a} and by extension, possibly supporting a wind origin
for the X-ray gas.  Theoretically, we should not be surprised to find
winds being driven off accretion disks \citep{king03a} and as it appears likely such
winds would extend to relatively close to the black hole \citep[e.g.][]{proga00a}, 
this motivates a
consideration of X-ray spectra in terms of absorption-dominated
models.  Indeed, X-ray spectroscopy directly detects absorption signatures indicative   
of an origin in a disk wind, as we now discuss.

\subsubsection{Parameter space for the X-ray absorbers}

X-ray observations show that absorbing gas exists 
over higher column
densities, and ionization-states than are measured in the UV band. 
Most local type\,1 AGN show numerous narrow absorption lines in the
soft X-ray regime where the effective area and spectral resolution of
various X-ray grating instruments are relatively high, and detected zones 
cover the full range of
ionization state to which observations are sensitive 
\citep[][and references therein]{blustin05a,mckernan07a,blustin07a}. 
 
Details of the  zones of absorbing gas, especially  for  
${\mathrm{N_H}} \la 10^{23}$cm$^{-2}$ are confirmed from grating
observations of absorption lines.  
At the low-ionization end, outflow velocities are typically 
in the range 100--2000\,km\,s$^{-1}$ (these are  usually consistent with 
having turbulent velocities of a similar order) and zones 
with $\log\xi < 0$ are commonly found 
\citep[e.g.][]{kaastra00a,kaspi02a,netzer03a, crenshaw03a}. 
Spectral modeling to date has assumed
the gas zones to be discrete, but a more realistic picture may be that
there is a wide range of ionization across a large-scale
outflow, with possibly multi-phase components present.
Some of the higher columns to date have 
been inferred from broad-band fits, as the opacity tends to be too
high to be able to detect discrete lines in soft-band grating
observations. An exception is the  highest ionization
zones that have $\log\xi \ga 3$, evident only from their telltale
absorption lines from Fe\,{\sc xxv} and Fe {\sc xxvi}; these lines have
equivalent widths of many tens of eV and those observed to date 
likely  arise from 
gas in the range $10^{23}- 5 \times 10^{24}$\,cm$^{-2}$
\citep[e.g][]{turner07a,turner08a} with outflow velocities 
ranging from  several thousands of km s$^{-1}$ up to relativistic flows. 
In many cases leakage of flux towards low energies is a strong indication
that at least some absorbing zones only partially cover the source,
with covering fractions suggested to be anything from a few tens of
percent to a full shell of gas.  An unambiguous signature of partial
covering absorption would be the detection of saturated line profiles
that have non-zero flux in the line core, as in UV spectroscopy of AGN:
unfortunately, as many of the X-ray absorption lines are 
not resolved even in grating data, line profiles 
cannot yet be used in this way. 

Spectral variability in type\,1 AGN is consistent with variations in
covering fraction of the gas. Detections of sources partially covered
by gas at the high end of this column density range may prove key to
settling the absorption/reflection debate 
(e.g. as in 1H\,0419--577, Turner et al. 2009).
Of course, absorbers that
are Compton-thick lead to an expectation of some accompanying
reflection from the same gas. As previously discussed, the level of
accompanying reflection or re-emission from the gas depends critically
on ionization-state and gas geometry.

\subsubsection{Reasonableness of partial covering as a concept}

In absorption models the covering fractions found for key gas zones
often come out $\la 1$ and variations of
tens of percent (with mean values often about $50\%$) 
may be used to explain observed spectral variability
\citep[e.g.][]{reeves02a,reeves03a,miller07a,turner07a}.

The most  common argument against partial-covering models is one based upon probability. The 
key point is that for an absorber to have some (possibly varying) covering fraction $\la 1$
it should contain structure on the same scale as the source it is obscuring.
If that structure arises because the absorber is composed of discrete clouds of some size,
then we should expect that size to be comparable to the source size.  Then, if the absorbing
clouds are distant from the source, there is some implied coincidence between the source
and absorber sizes, and furthermore to have a reasonable chance of seeing a partial-covering absorption
event there must be many such clouds distributed across the region.  Alternatively, the 
clouds might exist at radii comparable to that of the source.

A more likely scenario, however, is that the X-ray absorber exists as
part of a clumpy disk wind or as part of the broad-line region.  In
this case we would expect it to comprise a wide range of scale sizes,
perhaps with some power-spectrum of fluctuations as expected for
turbulent fluids, rather than being in the form of discrete clouds of
some fixed size.  Such distributions of clumpy absorbers are apparent
in disk wind models \citep[e.g.][]{proga08a}.  If the X-ray absorption
is in the form of a wind then the argument about probability is not
so relevant, an equatorial disk wind has a preferred plane and the
incidence of observation of rapidly varying partial-covering absorption depends on
viewing angle to the accretion disk.

A test of the reasonableness of absorption models may be possible by comparing emission-line
strengths with those expected from the absorbing gas (Section\,\ref{sec:linestrength}).  For
non-resonantly-scattered lines this may lead to an estimate of the fraction of the sky subtended by the wind
as seen from the source: if that fraction turns out to be extremely small that might argue against
the absorption model, or if it turns out to be some intermediate value that may instead allow
some estimate of wind geometry to be made.  If the ionization of the absorbing gas is sufficiently
high that the emission-line in question would be resonantly scattered, the problem becomes much
more difficult (Section\,\ref{sec:linestrength}).  We emphasize that the {\em absorber} model
parameters should be used to determine whether the material is sufficiently ionized that
a line could be resonantly scattered.  A line
observed in emission from apparently low ionization material may simply
originate in a different location.

In model fits, partial-covering models are indistinguishable from
`Thomson scattering' models: i.e. a source that appears to be  
partially covered cannot be distinguished from one where we are viewing  
a fully covered, absorbed continuum source plus some fraction of 
continuum that has been scattered energy-independently,
and this may provide an alternative explanation for the spectra of some sources. 

\subsubsection{Variability}

If absorption changes do produce the observed spectral variability
they must also produce variability in the observed broad-band X-ray
flux. The profiles of flux changes during AGN deep dips
(Section\,\ref{newabsorption}) may support this possibility.  In this
case, even if the continuum source is constant, we would expect to
observe large amplitude and rapid X-ray variability in type\,1 AGN and
this must be accounted for when interpreting the detailed
characteristics of X-ray variations.  It is not yet clear whether {\bf all} 
of the observed fluctuations are an artifact of such absorber
variations.

A corollary of the above is that if the variations in spectral shape
are explained as arising from partial-covering changes, then the
observation of systematic spectral flattening as a source dims implies
that the observed flux variations in such AGN are dominated by the
variable absorption and not by intrinsic continuum variations, as
these would not result in a correlation between brightness and
spectral shape.

\subsubsection{Location and geometry of the gas}

The distance of the absorbers from the ionizing source may be estimated from the value of ionization
parameter $\xi$ obtained from model fits to spectra.  Since
$\xi=L/r^2n$ (section~\ref{cthin}), then provided we can obtain estimates for
the other quantities we may make an estimate of the location of the absorber, $r$.
For an absorber of any substantial opacity, $\xi$ varies through the material
and it is usual to assume that the absorber has a discrete inner face,
and to define $r$ as being the distance of the inner face from the ionizing source
and $\xi$ the value of ionization parameter at that face.
The ionizing luminosity is usually fairly well constrained by the knowledge of
X-ray flux, power-law photon index $\Gamma$ and source cosmological redshift,
although assumptions have to be made both about the form of the continuum (e.g. power-law)
and its extrapolation outside the observed energy range.  The largest uncertainty
arises from lack of information about the gas density $n$, so estimates of
distance based on this method should always be parameterized in terms of the unknown
density \citep[e.g.][]{turner08a}.  A plausibility argument can be used to choose
a range for 
the most likely value of density and distance:  for example 
the constraint $\Delta r \simeq {\mathrm{N_H}}/n \la r$ might be applied in the case where the depth through the absorber 
is thought to be smaller than the distance of the absorber from the central source and this 
would then yield an approximate lower bound on $n$ and a corresponding approximate
upper bound on $r \la L/\xi {\mathrm{N_H}}$.

Analysts should also be aware of the 
degeneracies in model fits between column density ${\mathrm{N_H}}$, $\xi$ and $n$.  For example, 
for a given column density, an absorber with low $n$ must be proportionally larger along the
line-of-sight than one with high $n$.  In that case there may be substantial
inverse-square-law dilution of the radiation field through the gas, if the required
source depth is comparable to the distance from the source,
so that at the outer face $\xi$ may be much lower than in a higher-density,
smaller-depth, model.  This condition arises for 
${\mathrm{N_H}}/n \ga \sqrt{L/n\xi}$ or $n \la {\mathrm{N_H}}^2\xi/L$.  In this case the
mean $\xi$ though the material would be lower than in a high-density model, 
and a given dataset thus tends to
be best-fit by a degenerate family of models in which $\xi$ is anti-correlated with
$n$.  Appropriate choices of ionizing luminosity and density need to be made when
calculating the expected opacity with code such as {\sc xstar}.  Even then, such
code assumes the absorber to have invariant density, whereas this is unlikely to
be true in practice: for example, in a steady wind, density may fall off inversely as 
distance-squared, canceling out in the value of $\xi$ the inverse-square-law fall-off
in the illuminating radiation.  These considerations should always be born in mind when 
assessing model choice.

Based on such estimates, some Compton-thick 
partial-covering absorbers appear to lie within the BLR (e.g. Turner et al 2009) 
while lower column gas 
generally appears to exist outside the BLR \citep{blustin05a}, possibly
associated with the dusty torus. 
Other estimates of the location of the gas come from 
observation of column variations that imply origins typically on the inner edge of the BLR 
\citep[e.g.][]{puccetti04a}. However, taking an overview of the 
recent observational findings for this class of source it appears that the inferred absorption arguably 
should be considered in the context of outflow models rather than orbiting clouds and therefore 
we do not review  discussions in the literature  based upon the latter. 
 
Mass flow rates can be estimated from an assumed
volume filling factor, measured velocity and the fitted gas parameters.  
In a review of soft-band absorbers, \citet{blustin05a} found that for a sample of 
local AGN the median mass outflow 
rate  was about $0.3$\,M$_{\odot}$\,yr$^{-1}$ compared to a median mass accretion rate 
approximately an order of magnitude lower at about $0.04$\,M$_{\odot}$\,yr$^{-1}$: other 
studies suggest the outflow and accretion rates may be comparable \citep{chelouche05a,mckernan07a}. 
The kinetic luminosity in those flows was unimportant compared to the bolometric luminosity. 
As yet there are too few tightly constrained results from the high-velocity 
gas zones to draw many global conclusions. However, a recent 
analysis of PG 1211$+143$ \citep{pounds08a} found a P\,Cygni signature 
of high-velocity gas that implied an outflow with solid angle 
around $\pi$ steradians. The constraint on global covering allowed \citet{pounds08a} 
to estimate the outflow energetics and to conclude that this zone of gas makes an 
energetically significant contribution to feedback to the host galaxy. 

\section{Towards a complete picture of AGN X-ray emission}\label{complexmodels}

Whatever the origin of spectral curvature around 6\,keV, AGN carry unambiguous evidence, 
from grating data, for absorption covering  
a wide range of column, ionization-state and radial locations.  
When absorption models are considered to explain 
the observed systematic spectral and flux variability then the parameters 
implied for the gas suggest the material includes clumps of high column density with 
low covering fraction, that indicate that it would be a 
misleading oversimplification to continue to think in terms of 
individual single clouds.  The most natural gas geometry is likely 
an equatorial disk wind.  

Physically, we should expect that a black hole accreting at close to the Eddington limit should generate
significant outflows, and it seems likely that such an accretion disk would
have an atmosphere and that the atmosphere would be outflowing. 
 The column densities
required to significantly affect the observed X-ray spectra are not extreme compared with theoretical
expectation: estimates by 
\cite{king03a} indicate that sources with a high Eddington ratio would be expected to have 
Compton-thick winds 
outflowing at high velocity.  Early wind models for supercritical AGN had also 
suggested the 
winds would be optically thick in the central regions and that the continuum source would 
be hidden from direct observation \citep{camenzind83}. 
In many ways then, it is more incumbent upon modelers to justify why absorption can be ignored
in model fits rather than to justify why absorption should be included. 

Whether variable absorption can explain the full range of X-ray
spectral variability is more contentious, although SWIFT observations
of the brightest nearby AGN show that the X-ray emission from AGN is
more variable in those that are more absorbed \citep{beckmann07a},
implying that opacity variations play a significant role in
source variability.  It has been found that a model of variable  
partial covering for the variable X-ray spectrum of MCG--6-30-15 can
explain the spectral shape, including red wing and hard X-ray excess 
without requiring any inner disk reflection
\citep{miller08a}.  The model seems to require an anti-correlation
between illuminating luminosity and covering fraction, although the model fits are
not unique and other possible solutions exist.  If a systematic
dependence between covering fraction and luminosity is confirmed, this
may point to a picture in which the continuum source changes in size
systematically with luminosity, appearing and disappearing behind an
absorption structure such as a clumpy disk wind or atmosphere.
Changes in source size have previously been discussed in the context
of models of coronal plus disk emission \citep[e.g.][]{haardt97a}. 

\begin{figure}
\begin{center} 
\hbox{
\includegraphics[width=90mm,height=110mm,angle=-90]{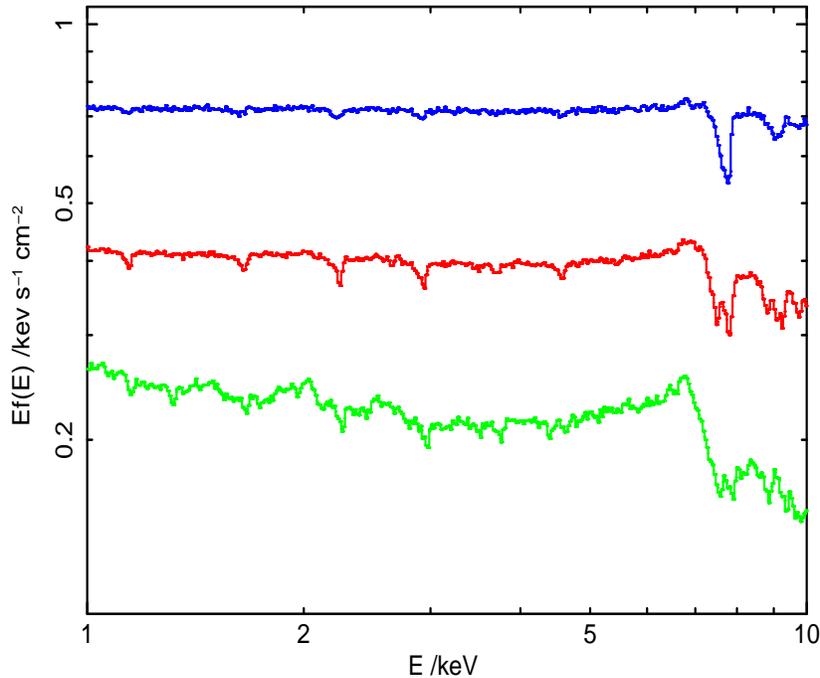}
}
\end{center}
\caption{\footnotesize
{Sample spectra computed for a viewing angle 60$^{\rm o}$ from the polar axis
for the equatorial smooth wind models of \citet{sim08a}.  System parameters
are chosen to be appropriate for the case of Mrk~766 with a wind originating at
radius 100\,\rg\ (see \citealt{sim08a}).
The wind opening half angle in this case is 45$^{\circ}$. 
From top to bottom are shown mass flow rates of 0.1, 0.3 and 1.0\,M$_{\odot} {\rm yr}^{-1}$.
The plotted spectra are all normalized to the input primary power-law spectrum
with photon index $\Gamma=2$.
}} 
\label{fig:simwind}
\end{figure}

In spectral fitting and in the above discussion, 
attempts to describe complex AGN X-ray spectra, beyond simple power-law models, 
have tended to be classed as either ``absorption-dominated'',
in which case a radiative transfer code such as {\sc xstar} \citep{kallman01a, kallman04a} 
is used to model a foreground absorber, or else as ``reflection-dominated'', where
the dominant spectral signatures are assumed to arise from a slab-like reflecting surface
modeled by e.g. {\sc reflion} \citep{ross05a}.  In practice it is likely that {\em both} 
absorption and reflection occur naturally, and in fact that both processes may arise simultaneously
in a clumpy wind from an accretion disk.  Such an environment is much more difficult to model 
because the distribution of absorbing/reflecting gas is completely unknown, but recent modelers
have attempted to predict the X-ray spectra of these more complex scenarios in order to make a comparison
with observation.  \citet{schurch07a} and \citet{schurch08b} have modeled the absorption component
expected from a disk wind by taking shells of {\sc xstar}-modeled absorbing zones (noting that
{\sc xstar} is strictly a 1D radial calculation of absorption in an assumed spherically symmetric
absorber and that it neglects scattering).  \citet{schurch08b}
have shown how even this simple approach, when applied to the disk wind simulations of 
\citet{proga04a}, can yield complex spectra that carry many features that
are observed in actual AGN.   
\citet{sim08a} have used a Monte Carlo method to calculate disk wind spectra, so far for
simple wind geometries, but making a more complete treatment of the radiative transfer that
includes the effects of scattering/reflection as well as absorption.  Their initial treatment
was limited to highly ionized material and to consideration of K shell ions, but further work
is extending the models to lower charge states and to more complex wind geometries.
The wind spectra produced by \citet{schurch08b} and \citet{sim08a} naturally explain the energy-shifted 
absorption lines 
observed in recent X-ray spectra, and \citet{sim08a} 
also show that  electron scattering, line scattering and recombination contribute to a 
significant emission of Fe K$\alpha$ with a pronounced red  wing, which 
may provide another possible explanation of the common observational signatures in AGN spectra. 
Earlier work on outflows by \citet{titarchuk03a} had also shown that 
significant 
line broadening would occur from Thomson scattering in outflows around 
black hole systems and that this scattering can produce broad features 
that match the observed data around 6 keV. 
Such scattering  effects are difficult 
to separate from strong relativistic effects  in current data. 

\section{Future prospects}\label{sec:prospects}

In the crucial energy range covering the Fe\,K-shell spectral features, 
current X-ray data offer either high throughput 
with modest spectral resolution, e.g. the CCDs of {\it XMM-Newton} and {\it Suzaku} 
with energy resolution FWHM\,$\simeq 130$\,eV at 6\,keV, or 
improved spectral resolution with low effective area, e.g. {\it Chandra} 
HEG having 
energy resolution FWHM\,$\simeq 40$\,eV  at 6\,keV. These data 
limitations have made it difficult to 
distinguish between absorption and reflection models for the curvature around 6\,keV. 
Successful flight of an X-ray calorimeter 
should resolve the issue.  The calorimeter design is currently being 
improved  compared to that used for the previous attempts with the {\it ASTRO-E} 
XRS-1 and  \suzaku\,XRS-2 calorimeters. The spectral resolution 
now achievable with calorimeters is FWHM\,$\simeq 7$\,eV, but 
resolution FWHM\,$\simeq 4$\,eV is projected for the calorimeters that should fly 
aboard {\it ASTRO-H} and FWHM\,$\simeq 2.5$\,eV for those on the 
{\it International X-ray Observatory} ({\it IXO}).  

\begin{figure}
\begin{center} 
\hbox{
\includegraphics[width=100mm,height=100mm,angle=0]{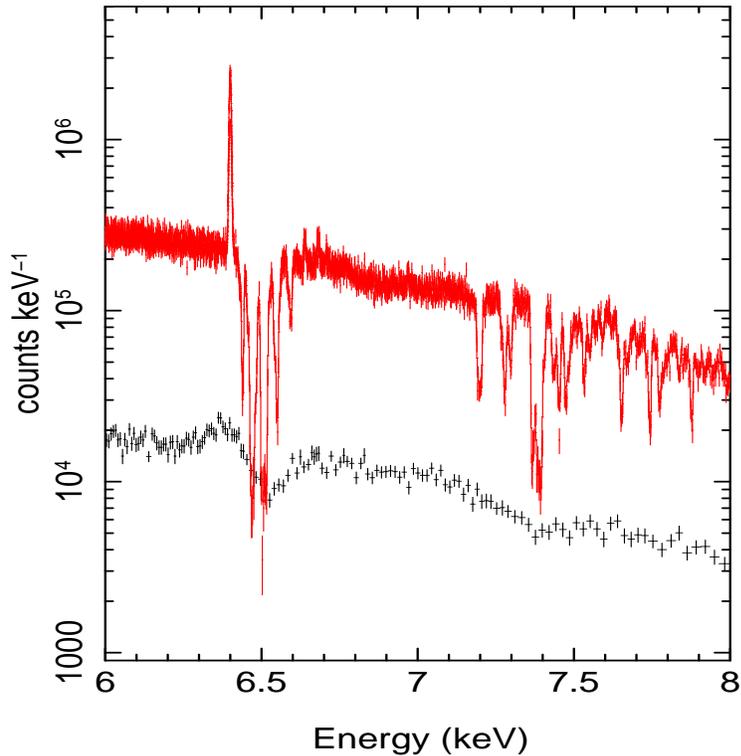}
}
\end{center}
\caption{\footnotesize
{A comparison of a model AGN spectrum as would be 
viewed by IXO (red) and \suzaku XIS0 plus XIS3 CCDs (black). 
Both spectra represent a 100\,ks exposure on a source of flux 
$F_{\mathrm{2-10 keV}} = 2.5 \times 10^{-11}$erg\,cm$^{-2}$\,s$^{-1}$  
having a power-law with photon index $\Gamma=2$ viewed
through an absorbing column with
$\mathrm{N_H} =5 \times 10^{23} {\mathrm{cm}^{-2}}$ and  $\log\xi=2.75$. 
The line emission is shown from the absorbing gas plus an additional Fe\,K$\alpha$ 
emission line from neutral material with equivalent width 100\,eV and line width\,$\sigma=3$\,eV. 
The simulation includes no velocity shift for either emitter or absorber.
}} 
\label{fig:comparison}
\end{figure}

{\it ASTRO-H} is planned for a launch around 2013 and, in addition to the calorimeter,
it will carry the first imaging hard-X-ray detector contributing to a
total bandpass covering 0.3--600\,keV.  
Closely following {\it ASTRO-H} will be the {\it Simbol-X} mission, jointly supported by the 
French (CNES) and Italian (ASI) space agencies and having a target launch date of 2014. 
{\it Simbol-X} will provide X-ray imaging up to 80 keV with approximately two orders of magnitude 
 improvement in 
sensitivity compared to the non-focusing instruments that have 
operated above 10\,keV so far. {\it ASTRO-H} and {\it Simbol-X} will  yield a 
wealth of valuable new information on Compton-thick absorbed systems in 
AGN. {\it IXO} will be a
collaborative mission involving ESA, JAXA and NASA, and is planned to
be launched sometime beyond 2018.  In addition to the calorimeter,
{\it IXO} will carry a wide field imager and a X-ray grating
spectrometer dispersed onto CCDs.  Possible additional instruments for
{\it IXO} include an X-ray polarimeter, a hard-band instrument or a
high-resolution timing instrument.  The effective area of {\it IXO} is
currently planned to be 10,000\,cm$^{2}$ at 6\,keV and thus {\it IXO} is
poised to offer a factor of 20 improvement over the best effective area offered
by current CCD instruments, and the same factor improvement over the
best resolution available (at 6\,keV). The scientific return however
should be even greater than these numbers indicate because the high
throughput and resolution will be available together.  With the likely
combination of calorimeter and high throughput, {\it IXO} observation
exposures of 10\,ks would be sufficient to detect key diagnostic
features such as the Fe\,K$\beta$ UTA in bright Seyfert galaxies; this
complex is predicted based on some absorption models for particular
AGN but is not detectable with current instruments
\citep[e.g.][]{miller08a}. Thus calorimeter observations with
reasonable exposure times would offer the possibility of resolving the
origin of the curvature around 6\,keV by isolating weak absorption 
features so that we can build up a clearer picture of the nuclear
regions.
Figure~\ref{fig:comparison} shows an example of how 
clearly spectral structure would be measured by a future calorimeter 
compared to data obtained for the same intrinsic spectrum and exposure time with \suzaku CCDs. 

Another area of current interest are the rapidly variable narrow lines:   
the  rapid variability in energy observed for these  energy-shifted  features 
means that one cannot simply 
make a long integration and obtain a stronger detection. Tracking the line changes 
for a sufficiently long time however, would allow the pattern of changes in these weak lines 
to be compared with predictions from hotspot and wind models 
and compared to random noise fluctuations.  
Changes in the flux and spectrum above 10\,keV, where the reflection continuum excess peaks,
are expected to be associated with line signatures having a disk hotspot origin.   Clearly, 
a hard-band detector would be invaluable for use in conjunction with a calorimeter. 
 We should be able to  confirm the line origin using 
time-resolved spectroscopy with {\it IXO}, because we could accumulate sufficient photons in azimuthal 
samples round the disk orbit to see what patterns emerge; we are currently limited to integrating much longer than 
the likely timescales of interest to obtain sufficient photons for spectral analysis.

%\begin{minipage}{100mm}
\begin{table}
\caption{Flux (2--10\,keV) and orbital timescale at 10\,\rg\ for
the Tartarus database sample of bright AGN.}
\label{tab:1}       % Give a unique label
\begin{tabular}{lrlr}
\hline\noalign{\smallskip}
Target & 2-10 keV & flux & T$_{\rm orb}$ \\
 & \multicolumn{2}{c}{/$10^{-11}\, {\rm erg\, s^{-1}\, cm^{-2}}$} & /ks \\
\noalign{\smallskip}\hline\noalign{\smallskip}
IC 4329A$^a$ & 7.0 & & 1.0 \\
MCG-6-30-15$^b$ & 4.0  & & 2.0 \\
NGC 4051 $^a$ & 2.0 & & 2.0 \\
NGC 5506$^c$ &  7.0 & & 2.0 \\
Mrk 766$^d$ & 2.0 & & 4.0 \\
Mrk 335$^e$ & 1.0 & & 5.1 \\
NGC 7314$^f$ & 4.0 & & 5.1 \\
NGC 7469$^e$ & 3.2 & & 7.1 \\
NGC 4593$^e$ & 4.5 & & 8.1 \\
NGC 4151$^a$ & 10.0 & & 13.2 \\
MCG+8-11-11$^g$ & 2.3 & & 15.2 \\
NGC 3516$^e$ & 5.0 & & 23.3 \\
NGC 3783$^a$ & 7.0 & & 29.3 \\
NGC 3227$^e$ & 2.8 & & 44.5 \\
NGC 2992$^e$ & 0.4 & & 52.6 \\
MCG--5-23-16$^g$ & 9.0 & & 70.8 \\
Mrk 509$^e$ & 6.6 & & 72.9 \\
Fairall 9$^e$ & 2.5 & & 82.0 \\
MR 2251-178$^h$ & 5.0 & & 99.2 \\
NGC 7213$^e$ & 3.0 & & 99.2 \\
Mrk 841$^e$ & 1.0 & & 101.2 \\
NGC 5548$^e$ & 5.0 & & 111.3 \\
Arp 102B$^i$ & 1.1 & & 141.7 \\
NGC 2110$^e$ & 3.5 & & 202.4 \\
MCG--2-58-22$^g$ & 3.3 & & 354.2 \\ 
\hline\noalign{\smallskip}
$^a$ \citet{peterson04a} &  & & \\
$^b$  \citet{wang04a} & & & \\
$^c$ \citet{bianchi03a} & & & \\ 
$^d$ \citet{wandel02a} & & & \\
$^e$ \citet{woo02a} & & & \\
$^f$ \citet{padovani88a} & & & \\ 
$^g$ \citet{bian03a} & & & \\
$^h$ \citet{morales02a} & &  & \\
$^i$ \citet{wu04a}  & & & \\
\end{tabular}
\end{table}
%\end{minipage}

Considering the case where lines are observed from hotspots orbiting the black hole,
the Keplerian orbital time varies with the mass of the central black hole as 
$t_{orb} \simeq (r/9\mathrm{r_g})^{3/2}$ M$_8$ days, where  M$_8$ is the central mass in units 
$10^8$ M${_{\rm \odot}}$. At $r=10$\,\rg\ the orbital timescale is just $1000$\,s 
for an AGN with a non-rotating (Schwarzschild metric) 10$^6$\,M$_\odot$ black hole, 
rising to 100\,ks for a source with a 10$^8$\,M$_\odot$ black hole;  
even for the latter case 
the timescale is approximately equal to the integration time required to detect the weak features of 
interest in an AGN spectrum. While there have been indications of line profile evolution 
matching that expected from an orbiting hotspot in a few tantalizing cases \citep{longinotti04a, turner06a} 
observational progress has been very limited.  
The detect-ability of hotspot  lines is a function of 
both the orbital timescale (and therefore black hole mass) 
and the source flux, as both quantities determine the 
accumulated photons in time-selected spectral bins that would reveal 
detectable profile evolution. 

\begin{figure}
\begin{center} 
\hbox{
\includegraphics[width=100mm,height=110mm,angle=-90]{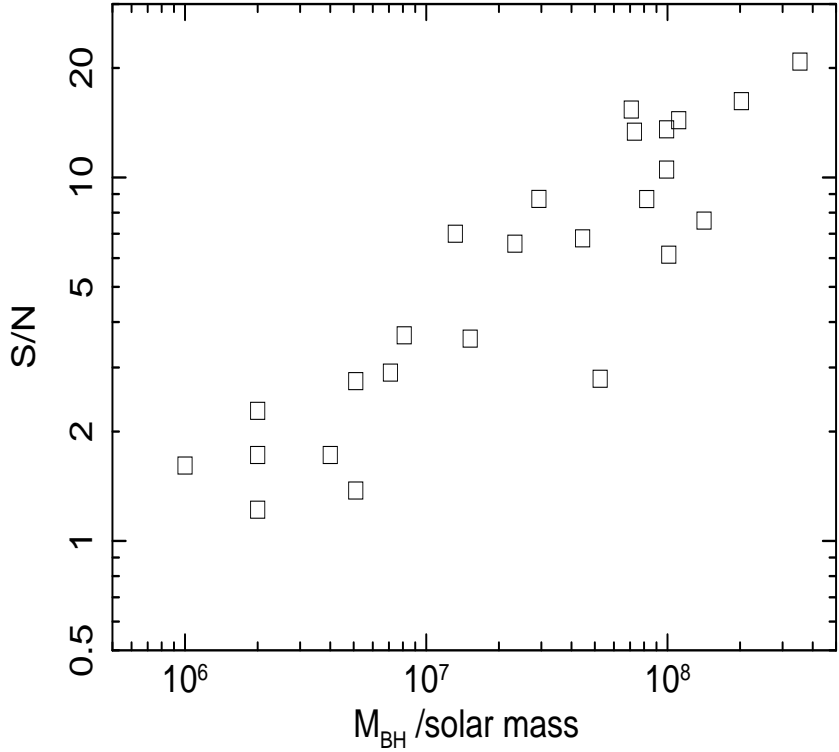}
}
\end{center}
\caption{\footnotesize
{The signal-to-noise that would be obtained for a line detection in 
one azimuthal portion, covering a 90 degree section of an orbit,
at emitting radius 10\,\rg\ plotted against the black hole mass for a number of bright AGN. 
Each plotted point represents a source listed in Table~\ref{tab:1}.  
}} 
\label{fig:merit}
\end{figure}

To track the evolution in line profile at 10\rg\ one might wish to take a minimum of 
four spectral samples around the disk orbit.  Figure~\ref{fig:merit} 
shows the signal-to-noise that would be obtained in an IXO 
exposure corresponding to one quarter of a disk orbit, as a function of 
black hole mass. For the simulation, a simple line was used to represent 
a sample of the profile falling on the red side of the rest energy, with the integration 
azimuthal section giving a line energy of 5.5\,keV. The line was conservatively 
taken to have an equivalent width 
of 30\,eV, and a width $\sigma=30 $\,eV. The simulation assumed an effective area 
of 9000\,cm$^{2}$ at 5.5\,keV for IXO. Points on the figure represent the 
same line seen at 10\rg\ in different sources comprising 
a sample of bright AGN (Table\,\ref{tab:1}), those
with flux in the 2--10\,keV band greater than $5 \times 10^{-11}$\,erg\,s$^{-1}$\,cm$^{-2}$
in the Tartarus\footnote{http://tartarus.gsfc.nasa.gov/}
compilation of AGN (having systematically reduced and fit  {\it ASCA} spectral results). 
It can be seen that many sources have timescales at 10\,\rg\ in the range of a few tens of ks, 
ideal for study with {\it IXO}. The plot can be used to assess potential observations using 
{\it ASTRO-H} if the values are scaled appropriately using the ratio of the effective areas. 
As noted previously, details of the line variability should yield key system parameters such as 
emitting radius and system inclination in the case of a hotspot origin; launch radius, 
acceleration/deceleration mechanism and mass loss rate if a wind origin were confirmed. 

\section{Conclusions}

Almost a decade of grating data from 
{\it Chandra} and {\it XMM-Newton} has helped shape our understanding of the circumnuclear environments in 
AGN.  The warm absorber generally appears as a multi-layered gas complex spanning a range of 
ionization parameter.  It is likely that this absorption is largely responsible for the
soft excesses seen in many AGN X-ray spectra and must be taken into account when fitting models
to spectra.  Higher columns of gas are also detected:
data from {\it XMM-Newton} CCDs, the {\it Chandra} HEG and the {\it Suzaku} XIS have  recently 
revealed  K-shell absorption lines from  Fe\,{\sc xxv} and Fe\,{\sc xxvi} ions in a number of AGN, tracing 
near Compton-thick gas.
In a number of AGN, such gas appears outflowing with
velocities from hundreds of km\,s$^{-1}$ to $\sim 0.3$\,c.  It is expected that disk
winds develop in high Eddington-ratio AGN, and such a phenomenon would be consistent
with the asymmetric UV emission lines seen in AGN.  
The most recent observations at energies above 10\,keV also indicate the presence of
high column densities of gas that partially cover the source and provide a natural
explanation for the hard-band excesses that are often observed.
Partial-covering by additional gas lying within the known range of column and ionization 
can explain observed X-ray spectral curvature and variability at lower energies, 2--8\,keV.
The key layers associated with such partial-covering solutions 
have not yet been unambiguously identified via narrow spectral features . However, 
occultation and de-occultation events also indicate the importance of dynamically variable 
absorbers.  
It now seems highly likely that, rather than having a direct view of a ``naked'' accretion disk 
dominated by a power-law continuum plus disk reflected emission, the view to the
central regions is in fact complex, with absorption covering a wide range of densities,
ionization and dynamics, likely coupled with reflection from material over a wide range of
radii.  With the recognition of the greater complexity of
AGN X-ray spectra, the inference of black hole spin from model fits is seen to be difficult
with the current generation of data.
However, there is much to learn about the accretion process from studying the complex inner regions
at radii $\la 100$\,${\mathrm{r_g}}$. 
New models are currently being developed to study the radiative transfer
through disk winds and new high resolution data, and imaging data in the hard X-ray band,
should provide crucial new insights in the coming decade.

\begin{acknowledgements}
The authors thank Stuart Sim, Ian George, Tahir Yaqoob and James Reeves for comments that significantly  
improved this manuscript. 
TJT acknowledges NASA grant NNX08AJ41G. LM acknowledges 
STFC grant number PP/E001114/1. 

\end{acknowledgements}

\bibliographystyle{spbasic}      % basic style, author-year citations
\bibliography{xrayreview_feb2}   % name your BibTeX data base

\end{document}